\documentclass[sigconf,table]{_acm/acmart}

\AtBeginDocument{%
  \providecommand\BibTeX{{%
    \normalfont B\kern-0.5em{\scshape i\kern-0.25em b}\kern-0.8em\TeX}}}

\copyrightyear{2025}
\acmYear{2025}
\setcopyright{cc}
\setcctype{by-nc-sa}
\acmConference[CHI '25]{CHI Conference on Human Factors in Computing Systems}{April 26-May 1, 2025}{Yokohama, Japan}
\acmBooktitle{CHI Conference on Human Factors in Computing Systems (CHI '25), April 26-May 1, 2025, Yokohama, Japan}
\acmDOI{10.1145/3706598.3714164}
\acmISBN{979-8-4007-1394-1/25/04}

\usepackage{lipsum}
\usepackage{subcaption}
\usepackage{macros}
\usepackage{float}
\usepackage{wrapfig}
\usepackage{placeins}

\newcommand{\etal}{et al.}

\begin{document}

\title[Malleable Overview-Detail Interfaces]{Malleable Overview-Detail Interfaces}

\author{Bryan Min}
\email{bdmin@ucsd.edu}
\orcid{0009-0003-0657-4398}
\affiliation{%
  \institution{University of California San Diego}
  \streetaddress{9500 Gilman Dr}
  \city{La Jolla}
  \state{California}
  \country{USA}
  \postcode{92093}
}
\author{Allen Chen}
\email{aschen@ucsd.edu}
\orcid{0009-0008-7961-8634}
\affiliation{%
  \institution{University of California San Diego}
  \streetaddress{9500 Gilman Dr}
  \city{La Jolla}
  \state{California}
  \country{USA}
  \postcode{92093}
}
\author{Yining Cao}
\email{rimacyn@ucsd.edu}
\orcid{0000-0002-3962-2830}
\affiliation{%
  \institution{University of California San Diego}
  \streetaddress{9500 Gilman Dr}
  \city{La Jolla}
  \state{California}
  \country{USA}
  \postcode{92093}
}
\author{Haijun Xia}
\email{haijunxia@ucsd.edu}
\orcid{0000-0002-9425-0881}
\affiliation{%
  \institution{University of California San Diego}
  \streetaddress{9500 Gilman Dr}
  \city{La Jolla}
  \state{California}
  \country{USA}
  \postcode{92093}
}

\renewcommand{\shortauthors}{Min et al.}

\begin{abstract}
The overview-detail design pattern, characterized by an overview of multiple items and a detailed view of a selected item, is ubiquitously implemented across software interfaces. Designers often try to account for all users, but ultimately these interfaces settle on a single form. For instance, an overview map may display hotel prices but omit other user-desired attributes. This research instead explores the malleable overview-detail interface, one that end-users can customize to address individual needs. Our content analysis of overview-detail interfaces uncovered three dimensions of variation: content, composition, and layout, enabling us to develop customization techniques along these dimensions. For content, we developed Fluid Attributes, a set of techniques enabling users to show and hide attributes between views and leverage AI to manipulate, reformat, and generate new attributes. For composition and layout, we provided solutions to compose multiple overviews and detail views and transform between various overview and overview-detail layouts. A user study on our techniques implemented in two design probes revealed that participants produced diverse customizations and unique usage patterns, highlighting the need and broad applicability for malleable overview-detail interfaces.

\end{abstract}




\begin{teaserfigure}
    \includegraphics[trim=0cm 0cm 0cm 0cm, clip=true, width=\textwidth]{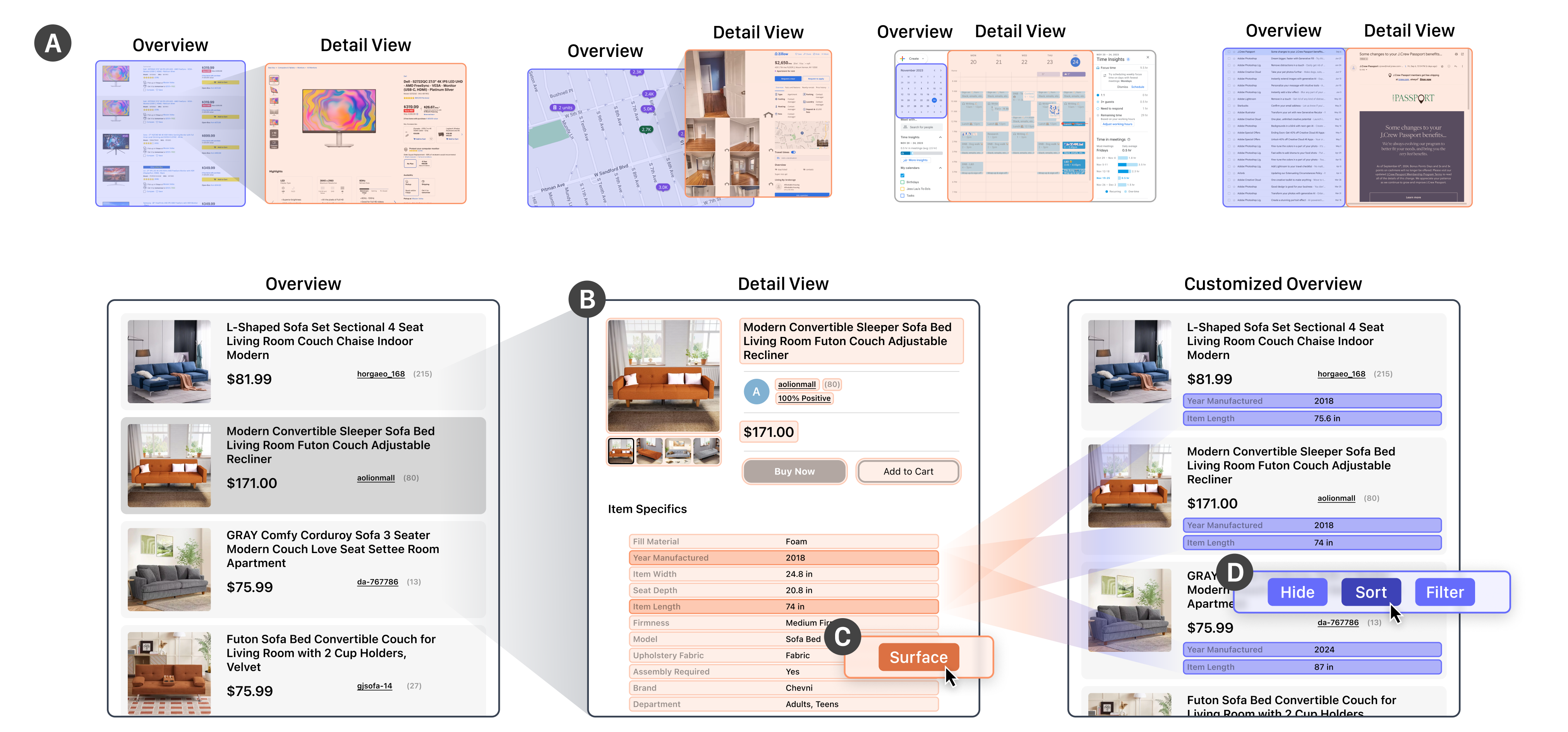}
    \caption{(A) We perform a content analysis of overview-detail interfaces (blue: overview; orange: detail view) to understand its variations in the wild, finding three key dimensions: \textit{content}, \textit{composition}, and \textit{layout}. (B) Based on these dimensions, we provide a set of interactions for end-users to customize their overview-detail interface. Among them, we contribute a set of techniques, \textit{Fluid Attributes}, enabling users to (C) surface attributes to the overview that only exist in the detail view and (D) operate on attributes they surface.}
    \label{fig:teaser}
    \Description{The figure depicts four images that encompass the primary contributions of this paper. The first image depicts three variations of overview-detail interfaces from our content analysis - highlighting key dimensions of content, compositions, and layout. The second image is taken from our design probe, showing the various interactions for end-users to customize the overview-detail interface to match their own needs and preferences. Among these interactions, we contribute a novel technique, Fluid Attributes, shown in the third and fourth images, enabling users to directly interact with and manipulate content attributes between the overview and the detail view. The third image shows, a user selecting attributes to be surfaced in the overview while the fourth image shows a user sorting the overview via a selected attribute.}
\end{teaserfigure}

\maketitle{}

\section{Introduction}
\label{section:introduction}

The overview-detail design pattern describes interfaces that typically consist of two views: an overview that presents a collection of items along with their key attributes and a detail view that provides an in-depth look at a selected item, including many, if not all, of its attributes \cite{Shneiderman1996OverviewFirst, tidwell1997pattern}. This pattern enables users to efficiently browse through a large collection of items and delve into specific ones of interest, facilitating one of the most essential operations in information tasks \cite{Norman1986ProgressiveDisclosure, Hornbaek2001OverviewDetailBenefits}. This is perhaps why the overview-detail pattern has become ubiquitous  throughout information systems, including shopping, booking, and rental sites, calendars, email clients, and numerous others (Fig. \ref{fig:teaser}A).

\begin{figure}
    \centering
    \includegraphics[width=1\linewidth]{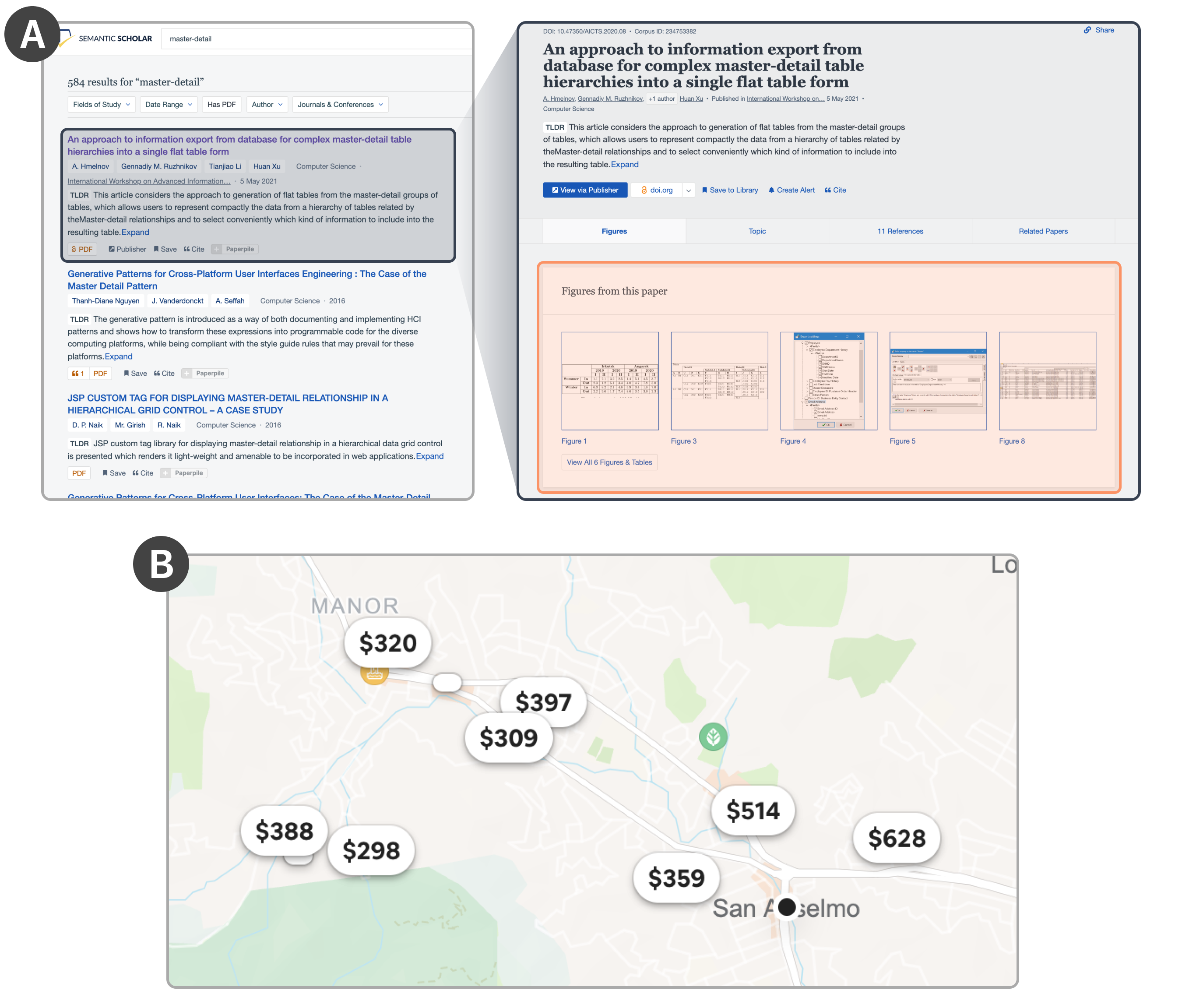}
    \caption{(A) Semantic Scholar includes a paper’s title, authors, abstract, and others, but does not include figures in the overview. (B) Airbnb displays pins with prices on a map, representing the nightly prices of each location. Users must click each pin to view details not shown in the overview, such as the paper figures, ratings, or number of bedrooms, and navigate through each full page of details.}
    \label{fig:md-limitations}
    \Description{This figure contains two website interfaces labeled (A) and (B). Interface A shows two screenshots of Semantic Scholar, an academic research platform.The left screenshot presents a list overview of search results showing the title, authors, abstract, and other details for research papers. The right screenshot displays the detailed view of a selected research paper. The user must open each paper’s detailed view to see figures and other specifics. The overview does not display these figures directly. Interface B shows a map with price tags pinned at different locations, representing the nightly prices of accommodations on Airbnb. If users want to view details like bedrooms, ratings, or reviews, they need to click on each pin and navigate to a full-page description of selected accommodation.}
    \vspace{-6pt}
\end{figure}

Typically, it is the designers and developers that determine how the overview-detail pattern is instantiated for a specific context, considering factors such as the layout and interaction between the overview and the detail view, as well as the different sets of attributes to present in these views.
While they strive to create the best design for their general user base, a single instantiation of the overview-detail pattern is inevitably not enough to match the diverse needs of all users and their usage contexts. 
For instance, separating the detail view to another page, such as in most shopping websites, can reduce clutter in the interface but incur repetitive view switching (Fig. \ref{fig:md-limitations}A). Meanwhile, displaying only the price of hotels on a map improves clarity but omits other attributes---such as ratings or the number of beds---that some users may prioritize (Fig. \ref{fig:md-limitations}B). Yet, few modern interfaces are \textit{malleable}, forcing the end-users to work with an interface set by their developers, regardless of the end-users' own needs and preferences. 

At its core, the overview-detail design pattern is intertwined with fundamental challenges in information foraging, sense-making, and decision-making \cite{coststructuresensemaking}.
First, the overview may not present the right abstraction of individual information entities to meet users' specific needs, forcing them to forage for details from other views \cite{infoforagingtheory}. 
Second, a system may hinder sense-making if the representation of the collection of items it provides does not adequately address users' needs \cite{zhang1994representations,Kirsh2010ThinkingWithExReps, Shneiderman1996OverviewFirst}.
Third, the spatial arrangement and interaction between multiple views may hinder necessary navigation among the different views.
Therefore, the key question we explore is: \textbf{instead of letting developers dictate how information is abstracted and presented, can we make the overview-detail design pattern malleable?} That is, can we empower end-users to flexibly customize overview-detail interfaces to form their own abstractions and spatial arrangements of information?

\noindent We break this goal into three research questions:

\begin{enumerate}[label=\textbf{[RQ\arabic*]}, leftmargin=*]
    \item  What is the \textbf{design space} of various implementations of this design pattern in real-world contexts,
and how can they inform the development of malleable overview-detail interfaces? 
    \item What are the \textbf{interface designs and interaction techniques} that can enable end-users to flexibly transform an overview-detail interface?
    \item To what extent do end-users \textbf{prefer and utilize} malleable overview-detail interfaces to achieve their information tasks? What are their novel \textbf{usage patterns}? Is the malleability we provide enough?
\end{enumerate}

We took the following steps to address these questions. 
First, we conducted a content analysis with a diverse sample of 303 instantiations of the overview-detail pattern and identified three key dimensions: \textit{content}, \textit{composition}, and \textit{layout} (\textbf{RQ1}).
Second, we developed interaction techniques that allow users to customize overview-detail interfaces along these three dimensions (\textbf{RQ2}).
For the content dimension, we developed a set of interactions, which we label as \textit{Fluid Attributes}, that enables users to surface new attributes from the detail view to the overview (Fig. \ref{fig:teaser}C), hide unwanted attributes in the overview, and use any selected attribute to filter and sort items (Fig. \ref{fig:teaser}D). Users can also prompt AI to perform these customizations automatically, reformat existing attributes, and create new ones.
For the composition and layout dimensions, we developed various options for end-users to select, such as setting the number of overviews in the interface and selecting layout options.
We demonstrate how our presented customizations could be designed and implemented through multiple applications.

Finally, we conducted a user study (N=12) to measure the utility and observe usage patterns of malleable overview-detail interfaces through two design probes in shopping and hotel booking tasks (\textbf{RQ3}).
We found that participants, given the same tasks, desired and produced diverse customizations along all three dimensions.
Furthermore, we discovered that participants' usage patterns resembled common tendencies in managing personal digital information and structuring tasks, demonstrating how malleable overview-detail interfaces integrate into broader, everyday workflows.

In summary, our work contributes:
\begin{enumerate}
    \item A content analysis of the overview-detail design pattern that reveals three key dimensions of the design space.
    \item Interaction techniques enabling end-users to customize overview-detail interfaces, offering diverse layouts and compositions of overviews and detail views, attribute manipulability across views, and AI-driven customizations.
    \item A user study demonstrating that participants effectively utilized malleable overview-detail interfaces to address diverse needs and revealing emergent usage patterns that inform design suggestions for future implementations.
\end{enumerate}

Our vision is to make all aspects of interfaces malleable.
We approach this vision by tackling \textit{one design pattern at a time}—analyzing variations of a single design pattern and exploring generalizable customization techniques that extend beyond their originally intended contexts.
We hope this paper not only encourages future interfaces to implement malleable overview-detail interfaces, but also for future research to adopt this paper's approach to explore how more design patterns can become malleable.

\section{Related Work}
\label{section:related_work}

\subsection{The Overview-Detail Design Pattern}

The overview-detail design pattern describes interfaces that provide two views: an overview that contains a collection of items and a detail view that provides full details of a selected item \cite{Shneiderman1996OverviewFirst}. This pattern, also referred to as ``master-detail’’ \cite{Seffah2015PatternsOH, Molina2002JustUI, Nguyen2012MDAndroid} and ``director-detail’’ \cite{Pastor2007ModelDrivenArchitecture}, visualizes information for users to view a large collection of items first before uncovering their details.
If overview-detail interfaces are designed with progressive disclosure in mind, they can offer a more approachable interface while still allowing users to access additional details \cite{Norman1986ProgressiveDisclosure, Shneiderman1996OverviewFirst}.

Prior research explored several variations of the overview-detail pattern in user interfaces. Case studies have explored how overview-detail interfaces can be nested to create hierarchical views \cite{Molina2002JustUI, Seffah2015PatternsOH}, how several detail views can be used simultaneously \cite{Saidi2016SplittingDetailViews}, and how different screen sizes may necessitate varying arrangements of the overview and detail views  \cite{Seffah2015PatternsOH, Nguyen2012MDAndroid}. 
Cockburn \etal{} have also explored how overviews and detail views can be arranged horizontally, vertically, and even along the ``z-plane'' by layering nested overviews and detail views along the z-axis, navigable by zooming. This z-axis arrangement also describes the structure of zoomable user interfaces such as Pad++ \cite{Bederson1994Padpp, Cockburn2009OverviewDetailReview}. 
A relevant but distinct design pattern also explored is the ``focus+context'' interface \cite{FirstFocusContextInterface, kosara2002focuscontextliterally}, where focused content is placed seamlessly with the rest, oftentimes by reducing the view of the context space by scaling down items \cite{bederson2000fisheye}, skewing the views in 3D \cite{Mackinlay1991PerspectiveWallFocusContext}, and reducing the display's resolution \cite{Baudisch2001FocusPlusContext}.  

These works demonstrate the diverse implementations of the overview-detail design pattern that researchers have developed, resulting in unique design choices tailored to specific contexts. However, our research aims to explore designs that enable users to flexibly customize the overview-detail interface according to their individual needs.

\subsection{The Need for Malleable Overview-Detail Interfaces in Information Tasks}

Many information tasks, such as foraging \cite{infoforagingtheory}, sensemaking \cite{coststructuresensemaking}, and decision-making \cite{mesh}, involve an iterative process of searching, browsing, collecting information across various sources, making sense of collected information through various representations, and collecting new information based on new insights \cite{coststructuresensemaking, Pirolli2005SensemakingProcess}. Throughout this process, users need representations that help them effectively organize and preview a large number of information entities and quickly determine relevant ones by inspecting their details. The overview-detail pattern directly responds to such needs and, therefore, is ubiquitous in information systems, including information visualization systems \cite{Card1999VisionToThink, bach2016TellingStoriesGraphicComics} and academic literature review tools \cite{Relatedly, Kang2023Synergi}.

As mentioned earlier, a single instantiation determined by developers is not enough. Therefore, many research systems that have been created to support information tasks are essentially providing additional overview-detail instantiations to compensate what could not be supported with existing ones. For example, prior work has explored comparison tables as a means to better organize decision-making criteria \cite{mesh}, hierarchical organizations of collected snippets to facilitate information management and re-finding \cite{Kang2022Threddy, Kang2023Synergi, fuse}, spatial canvases that scaffold the user's search and synthesis processes \cite{Ramos2022ForSense, Tashman2011LiquidText, Srishti2022InterWeave}, and node-link diagrams to contextualize users with their past browsing activities \cite{graphhistory}.

Another line of work, recognizing that a single instantiation is not enough, proposes providing multiple instantiations, such as those that enable multiple levels of details or transforming information representations. Pad++ introduced semantic zooming, which offers multiple levels of detail based on different zoom levels, providing an alternative to geometric scaling techniques used in traditional zoomable interfaces. Recent research has also built on this idea to generate these different levels of detail with AI \cite{Dang2022BeyongTextGeneration, Suh2023Sensecape, suh2024luminate}.
Victor similarly illustrated multiple representations through the ladder of abstraction, demonstrating how users can move between representations of various details to gain a more comprehensive understanding of a system \cite{Victor2011LadderOfAbstraction}. 
WritLarge, on the other hand, demonstrated techniques for moving between multiple axes of representation—semantic, structural, and temporal—on a digital whiteboard \cite{Xia2017WritLarge}. 
Graphologue also explored multiple representations, generating both text and node-link diagrams from an LLM in real-time to facilitate information comprehension \cite{Graphologue}.
We share the spirit of these systems that aim to provide multiple representations by providing end-users the ability to transform between different variations of overview-detail interfaces.

\subsection{Designing End-User Customizable Information Systems}

We share the research community's long-standing vision of creating malleable information systems for end-users to easily and expressively customize them to suit their preferences and information needs.
One research theme explores how to design the architecture of information systems to make them inherently customizable. For example, Haystack proposed a unified data model and a customizable interface layer to enable users to integrate diverse information sources and customize what information is presented to enable personalized information management \cite{adar1999haystack, Karger2007BDMHaystack}. Smalltalk focused on enabling users to directly modify the appearance and behavior of any object in the GUI \cite{kay1993smalltalk}. HyperCard explored enabling users to create dynamic and interactive content through a scripting language and graphical user interfaces \cite{Hypercard2024}. Finally, Webstrates treated web documents as dynamic media, enabling users to create custom representations of the same DOM element while facilitating real-time, dynamic collaboration \cite{Klokmose2015Webstrates}.

Another line of work explores the extension of existing systems to make them end-user customizable, particularly with web systems thanks to their accessible DOMs.
For instance, research has explored enabling end-users to directly customize a web interface by modifying the CSS attributes for website prototyping \cite{Kim2022Stylette, Visbug2024, Tanner2019Poirot}, embedding annotations between DOM elements to facilitate in-situ note-taking \cite{Romat2019SpaceInk}, and filtering, sorting, and making spreadsheet computations to personalize lists of items shown on a webpage \cite{Litt2020Wildcard, Huynh2006Sifter}. Research has also explored the potential of web mashups and automations enabled by direct manipulation and programming-by-demonstration techniques to automate workflows and reduce context-switching  \cite{Ghiani2016WebMashup, Bolin2005Chickenfoot, Tan2004Wincuts, min2023masonview, Hartman2007Dmix, Zhang2018Fusion, Lin2009Vegemite}.

Most of the above work, however, focus on exploring the underlying information architecture or the technical approach to enable customization for specific activities, applications, and platforms, rather than concretely and conceptually scoping how broadly generalizable their customizations are.
For example, it is not readily obvious how approaches taken by personal information management systems (e.g., Haystack \cite{Karger2007BDMHaystack}, Notion \cite{Notion2024}) should be generalized to shopping websites.
Moreover, it is unclear how systems like Sifter \cite{Huynh2006Sifter}, which recognize a common DOM structure for list views, can generalize to other list variations or even extend to different DOM structures.
In contrast, our research focuses on investigating the overview-detail design pattern that is ubiquitous across diverse information systems and platforms. The design space of the overview-detail pattern that we propose allows interface designers to easily recognize the variations of overview-detail interfaces that exist in the wild and also serve as a comprehensive guide of what all the possible customizations any overview-detail interface variation could theoretically support.

\subsection{Manipulable Attributes in Software Interfaces}

We build upon extensive work that has explored how manipulable attributes can be utilized in a variety of domains.
First, data visualization focuses on encoding data into visuals, and manipulating data attributes has been a core technique that enables users to create, customize, and interact with data visualizations \cite{liu2018dataillustrator, DataInk}.
Researchers have also used manipulable attributes to support creative tasks, enabling users to compose new graphical styles \cite{ObjectOrientedDrawing}, search for related graphical objects based on attributes \cite{Xia2017CollectionObjects}, and make color attributes manipulable across multiple representations \cite{Chevalier2012Histomages}.
Personal information management systems such as Haystack \cite{Karger2005HaystackSystem} have enabled users to manipulate attributes to modify how their personal information is presented, and database- and table-oriented software such as MS Access \cite{MSAccess2024}, Filemaker Pro \cite{FileMakerPro2024}, Notion \cite{Notion2024}, and Airtable \cite{Airtable2024} have also allowed the end-user to show and hide attributes on their interface.
Additionally, our content analysis of overview-detail interfaces has found e-commerce sites that allow end-users to show or hide a fixed set of attributes (Section \ref{section:existing-customizability}).

We see the overview and the detail view as two different containers of different sets of attributes, and by making both the attributes and views accessible and manipulable, we can enable end-users to flexibly customize them.
Our formulation of the malleable overview-detail design pattern acts as an amplifier, allowing us to leverage all prior techniques for making attributes manipulable for the overview-detail pattern and making them generalizable beyond each of their originally intended scenarios and contexts.
Additionally, we explore how AI can facilitate end-users in manipulating attributes within the overview-detail interface to achieve their diverse information needs, enabling users to prompt the interface to surface, hide, sort by, filter by, and select values in attributes.

\section{Content Analysis}
\label{section:content_analysis}

To better understand the various implementations of the overview-detail design pattern, we conducted a content analysis of overview-detail interfaces in desktop websites. Our content analysis was driven by the following questions:

\begin{enumerate}[label=\textbf{[Q\arabic*]}, leftmargin=*]
    \item What kinds of overview-detail interface arrangements exist in the wild?
    \item What types of information are presented in the overview and detail view, respectively?
    \item What aspects of overview-detail interfaces, if any, are already customizable by end-users?
\end{enumerate}

We present our findings with a design space of three dimensions of variations in overview-detail interfaces (\textbf{Q1}, \textbf{Q2}) and report existing customizability options found in our collection (\textbf{Q3}).

\begin{figure*}
    \centering
    \includegraphics[width=1\linewidth]{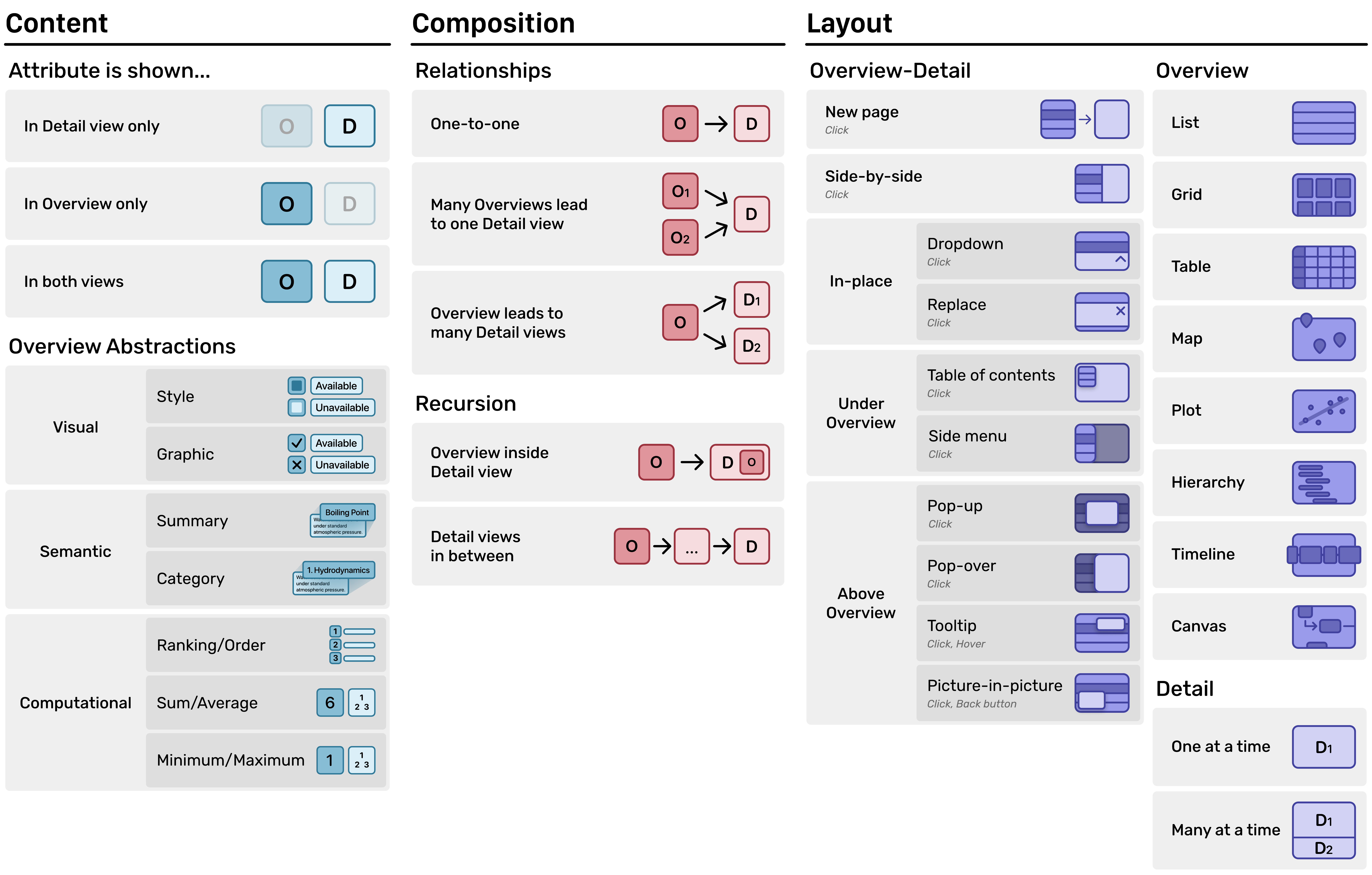}
    \caption{Design space of overview-detail interfaces on the Web: Content, Composition, and Layout.}
    \label{fig:design-space}
    \Description{This image provides a detailed breakdown of the design space of overview-detail interfaces used on the web, broken into three categories: Content, Composition, and Layout. Here's the description of each section: First, we begin with the Content dimension. The following subsections describe where a particular attribute is shown: In Detail view only: Represents scenarios where certain attributes are only visible in the detailed view. In Overview only: Attributes visible solely in the overview. In both views: Attributes visible in both overview and detail views. There are also several Overview Abstractions: Visual: Attributes like Style and Graphic can either be available or unavailable. Semantic: Attributes related to meanings like Summary or Category can be abstracted, such as a rating summary or topic labels. Computational: Attributes like Ranking/Order, Sum/Average, or Minimum/Maximum are shown through numerical abstractions. Next, we move onto the composition dimension. These Composed relationships include the following; One-to-one: One overview leads to one detail view. Many-to-one: Multiple overview points lead to one detail view. One-to-many: A single overview leads to multiple detail views. Composition also covers the following recursion: Overview inside Detail view: A recursive structure where an overview is composed with the detailed view. Detail views in between: Multiple detail views between overviews, often in a preview or hierarchical informational structure. Finally, we also have our Layout dimension. The Overview-Detail Layouts can New page: Clicking the overview opens the detailed view in a new page. Side-by-side: Overview and detail views appear next to each other. In-place: A dropdown or a replacement of content occurs in the same space. Under Overview: Detail appears under the overview via table of contents or side menu. Above Overview: Detail appears via pop-up, pop-over, tooltip, or picture-in-picture. The Overview itself can be displayed as a List, Grid, Table, Spatial map, Hierarchy, or Timeline. Lastly, the Detail Layout is highly variable, but can either show one view at a time or many at a time.}
\end{figure*}

\subsection{Methods}

\subsubsection{Data Collection}

The goal of our data collection was to collect a diverse and representative set of overview-detail interfaces that people regularly use. We initially collected the top 50 highest traffic websites in the United States based on data reported by Similarweb\footnote{\href{http://similarweb.com/}{http://similarweb.com/}. Accessed June 2024.}, a web analytics platform. However, as we analyzed this collection, we realized that many common websites were not included, such as flight booking or restaurant websites. To diversify our collection, we expanded to three more sources: 1) top 5 highest traffic websites in each of 33 web service categories\footnote{Names of 33 categories listed by Semrush in Appendix \ref{appendix:semrush-categories}.} listed by Semrush\footnote{\href{https://www.semrush.com/website/top/united-states/all/}{https://www.semrush.com/website/top/united-states/all/}. Data last updated May 2024. Accessed July 2024.}, another web analytics platform, 2) 20 randomly sampled websites from Wordpress’ showcase page to observe non-corporate websites such as blogs\footnote{ \href{https://wordpress.org/showcase/archives/}{https://wordpress.org/showcase/archives/}. Accessed July 2024.}, and 3) 20 randomly sampled websites from Awwwards’ gallery of ``2024 Sites of the Year'' to observe favorite designs from the UI/UX design community\footnote{ \href{https://www.awwwards.com/websites/sites_of_the_year/}{https://www.awwwards.com/websites/sites\_of\_the\_year/}. Accessed July 2024.}. After filtering our list (e.g., removing duplicates), our final sample consisted of 156 websites.
The first author examined all 156 websites, identifying 303 instances of overview-detail interfaces. Some websites featured multiple overviews linked to one detail view, or vice versa. For coding purposes, we defined each instance as a single overview-to-detail relationship. Thus, interfaces with multiple connections were recorded as separate instances.

\subsubsection{Coding Procedure}

We developed a codebook of overview-detail interfaces through an iterative open coding process involving collaboration among the first three authors. In the first round of open coding, we randomly selected 20 overview-detail instances from our dataset of 303, and the first three authors individually analyzed all 20 instances. The coders discussed and clustered their categories together to develop an initial codebook. In the next round, the first two authors independently coded a second sample of 20 instances, achieving an inter-rater reliability of 94.8\% agreement between the codes, measured by Cohen’s kappa. They discussed the remaining differences and revised the dimensions to reach a consensus. Given the high agreement rate, the authors determined that it was sufficient for the second author to complete the remaining coding, continuing to code in random samples of 20. After each round, the first two authors reviewed and discussed new potential dimensions and variations, revising the codebook as necessary. This process was repeated four times, stopping when no new dimensions or variations were found in the last three stages. In total, the authors analyzed 120 overview-detail instances.

\subsection{Design Space}

Our design space covers the following three dimensions:
\begin{itemize}
\item \textit{\textbf{Content}} describes whether the information is presented in the overview, the detail view, or both, and how the attributes in the overview are abstracted from those in the detail view.
\item \textit{\textbf{Composition}} describes how multiple overviews and detail views are connected together.
\item \textit{\textbf{Layout}} describes the spatial arrangement of overview and detail views within the interface.
\end{itemize}

We describe notable cases found in these three dimensions and present the full design space in Figure \ref{fig:design-space}.

\subsubsection{Content}

\begin{figure}
\vspace{4pt}
    \centering
    \includegraphics[width=1\linewidth]{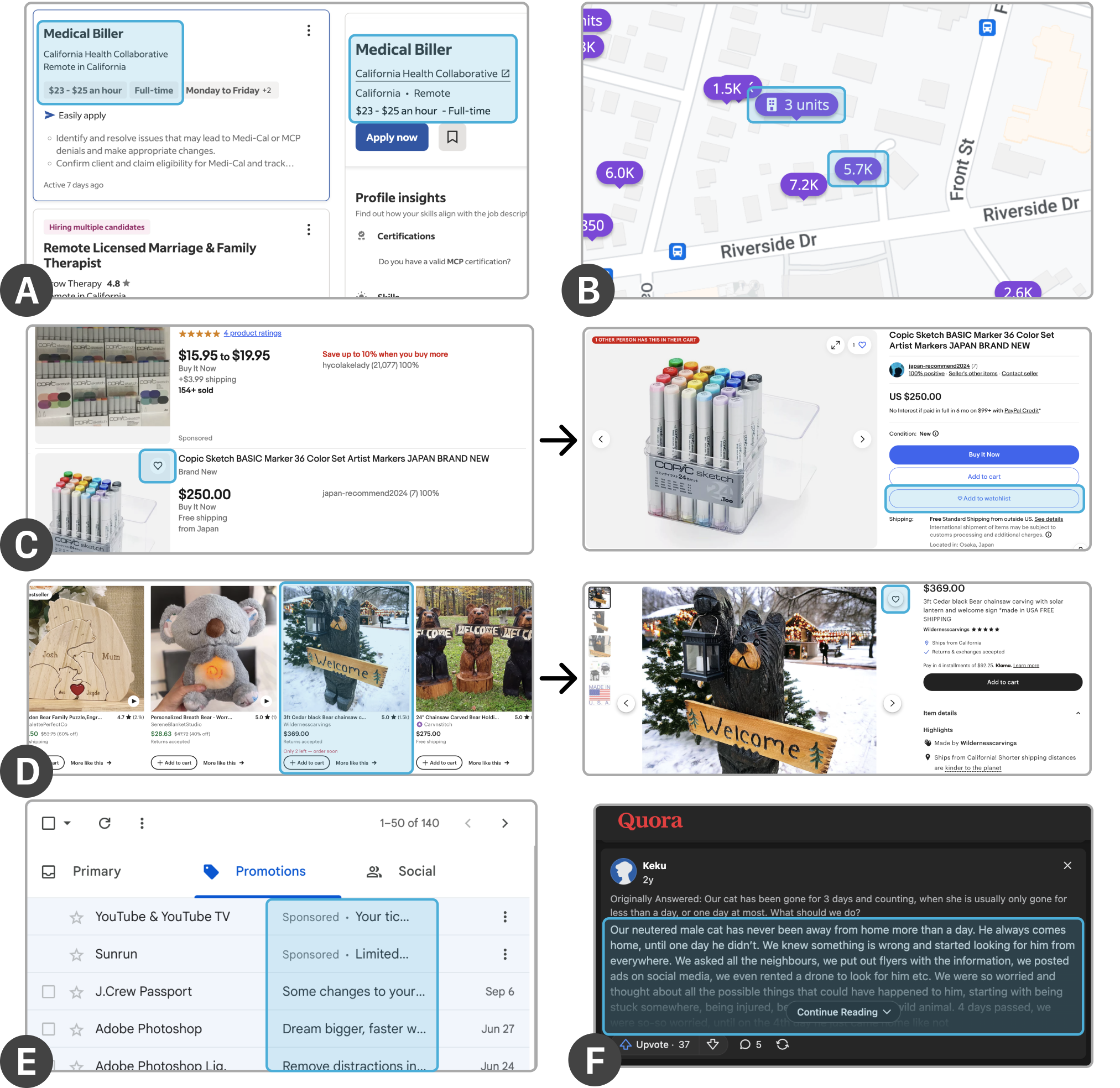}
    \caption{Example variations in the content dimension. (A) Indeed displays key attributes in both the overview and detail view. (B) Zillow's map shows prices, but if multiple units are present, it instead shows the number of units. (C) eBay provides the ``Add to Watchlist'' button in both views, whereas (D) Etsy provides it only in the detail. (E) Gmail cuts off long text with ellipses while (F) Quora does so with a gradient.}
    \label{fig:content-dimension}

    \Description{This image contains six panels labeled A through F, each demonstrating variations in how content is handled in overview-detail interfaces across different websites or apps: A) An image of Indeed's job listings interface: both the overview and detail\textbf{ }view display attributes of job title, company, location, and salary. B) A Zillow map showing properties. Depending on the number of units available, the overview displays different attributes like the price or number of units available. C) EBay's "Add to Watchlist" button is available in both the overview and detail views of product listings. D) Unlike eBay, Etsy shows the "Add to Favorites" option only in the detail view, not in the overview of product listings. E) Long blocks of text in Gmail's overview are truncated with ellipses (three dots), so the user has to open the detail view to see the full message. F) Here, long text is truncated in Quora's overview using a fade-out gradient instead of ellipses, prompting users to click to read more. Each panel shows different strategies for handling content visibility across overview and detail interfaces, highlighting where users see certain attributes.}
\vspace{4pt}
\end{figure}

Throughout our content analysis, we realized that some attributes were displayed only in the overview, some were displayed only in the detail view, and some were displayed in both. For example, the search results page for job openings on Indeed showed job titles, company names, and salary ranges in both views, while the overview omitted attributes like qualifications and benefits (Fig. \ref{fig:content-dimension}A).
Although we initially coded attributes as static information elements, we also observed this pattern in interactive elements such as `Save' buttons, text fields, and menus, with some sites including them in both the overview and detail views (Fig. \ref{fig:content-dimension}C), while others displaying them only in the detail view (Fig. \ref{fig:content-dimension}D).
Moving forward, we considered interactive elements to be attributes as well.
We also found that most rental and housing platform overviews presented prices for each place as pins on a map, but in Zillow, if a place had more than one unit available, the pin displayed the number of units available instead (Fig. \ref{fig:content-dimension}B).
In text-heavy overview-detail interfaces such as messaging apps and online discussion forums, we noticed long blocks of text in the overview were either cut short via ellipses (Fig. \ref{fig:content-dimension}E) or a fade-out gradient (Fig. \ref{fig:content-dimension}F).

We further observed that some attributes in the overview served as an abstraction of information in the detail view, subsequently identifying three forms.
First, we found visual abstractions, which were encoded into the style and visuals of the item. For instance, Ticketmaster used color shading to indicate the number of seats available per section, with sections that had more seats shown in a saturated blue, while sections with fewer seats appeared in lighter shades.
Second, we found semantic abstractions that represented the contents in the detail view with summaries or categories. For instance, ChatGPT's conversation interface generated a title that represented each conversation.
Lastly, we found computational abstractions that computed attributes from the detail view into a value. For instance, Southwest's website presents a calendar view of the cheapest flight prices for each day.

\begin{figure}
\vspace{8pt}
    \centering
    \includegraphics[width=1\linewidth]{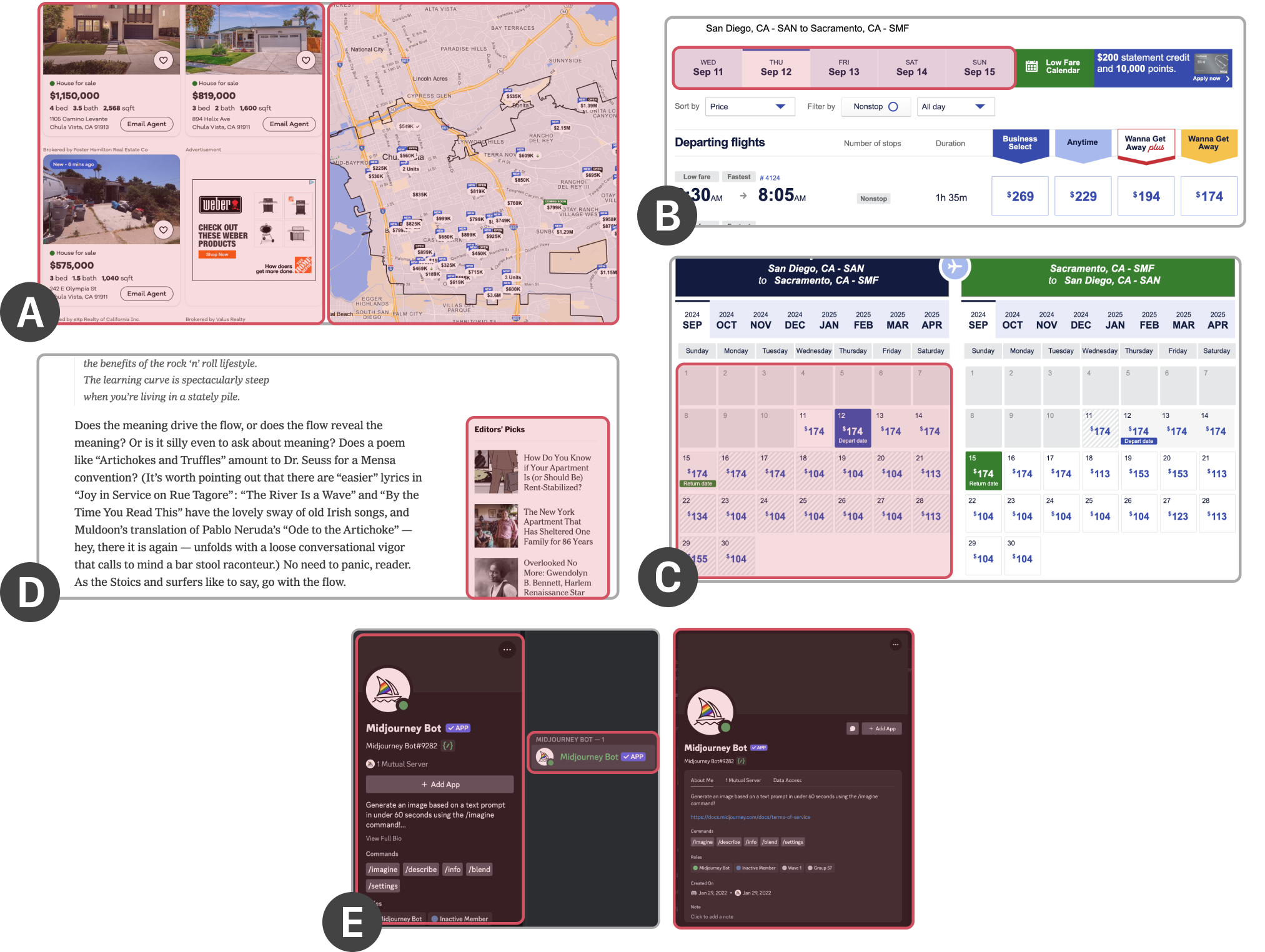}
    \caption{Example variations of the composition dimension. (A) Zillow shows a view composed of both a grid and a map. Southwest presents flight information through both (B) a 5-day table view and (C) a 30-day calendar view. (D) New York Times nests an overview inside a detail view with suggested articles. (F) Discord first shows a tooltip detail view of the user profile, and upon expanding, shows the full detail view.}
    \label{fig:composition-dimension}
\Description{The figure shows various examples of user interface compositions from different websites and applications. It's labeled A through E, each representing a different interface example: A) Zillow shows a real estate listing interface with two parts - a grid view on the left showing property details and images, and a map view on the right displaying the locations. B) Southwest Airlines shows a flight booking interface with a 5-day table view of flight options, prices, and times. C) Also Southwest Airlines displays a 30-day calendar view of flight prices, allowing users to see price trends over a month. D) New York Times shows an article view with suggested related articles nested within it. E) Discord displays a user profile interface, showing how a tooltip with intermediate details can be expanded into a full detail view. The image is meant to illustrate different ways that interfaces can be composed to show varying levels of detail and different views of information.}
\end{figure}

\subsubsection{Composition}

Additionally, we identified several overview-detail interfaces where multiple instances were composed together.

First, we found overview-detail relationships beyond just one-to-one. Many-to-one relationships presented multiple overviews such as a list overview and a map overview presented side-by-side (Fig. \ref{fig:composition-dimension}A), as well as different scopes such as a five-day list (Fig. \ref{fig:composition-dimension}B) and a 30-day calendar (Fig. \ref{fig:composition-dimension}C). We also found instances that synchronized navigation across the multiple overviews, where selecting an item in one overview would navigate users to the corresponding item in the other.

Second, we found instances of recursion---overview-detail interfaces nested within each other. 
For instance, additional overviews such as a list of articles appeared in the detail view of New York Times (Fig. \ref{fig:composition-dimension}D). 
This finding aligns with prior research that recognizes the variation of nested overview-detail views in hierarchical menu interfaces \cite{Seffah2015PatternsOH, Cockburn2009OverviewDetailReview}.
We also observed cases of ``intermediate'' detail views, in which a link or button inside one detail view opened another detail view containing more or all attributes of the item. For instance, clicking on a user in a Discord server's list of users opened a tooltip profile card, and clicking on the profile picture in the tooltip opened a full-screen popup, revealing further details about mutual friends and servers (Fig. \ref{fig:composition-dimension}E).

\subsubsection{Layout}

\begin{figure}
\vspace{4pt}
    \centering
    \includegraphics[width=1\linewidth]{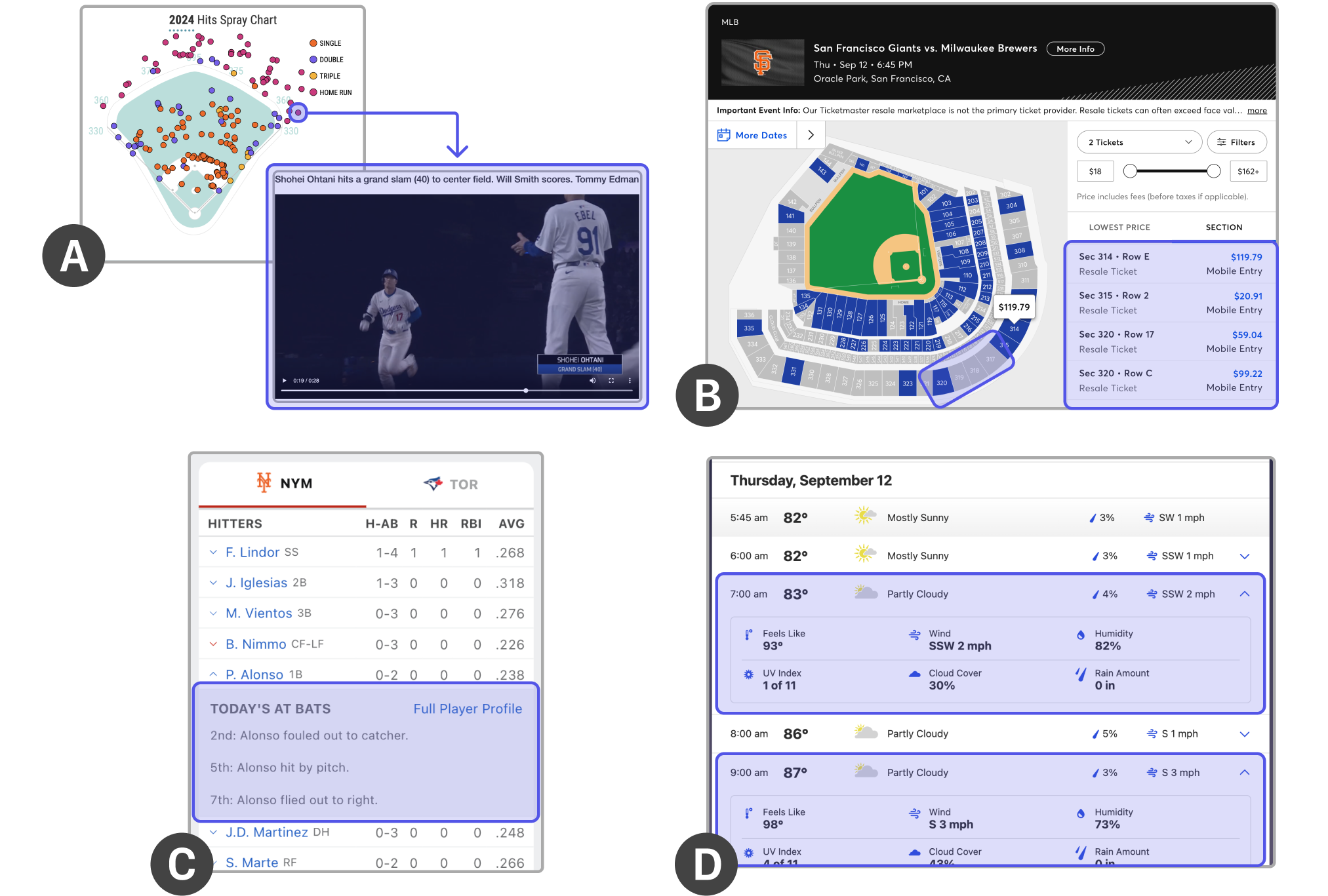}
    \caption{Example variations in the layout dimension. (A) MLB’s player profile page maps batting points onto the baseball field; clicking a plot point opens a detail view of a video replay of the point. (B) Ticketmaster’s seating selection page also shows a map overview; when a particular seating section is selected, the overview list changes to match the corresponding section. (C) ESPN shows each player's stats in a game, which can be opened as a dropdown. (D) Weather.com uses a similar dropdown layout, but unlike ESPN, multiple detail views can be opened simultaneously.}
    \label{fig:layout-dimension}

\Description{This image illustrates different variations in layout design for various sports and weather-related websites. The figure is divided into four sections, labeled A through D: A) MLB (Major League Baseball) Player Profile shows a baseball field diagram with colored dots representing batting points. An arrow points to a larger image, which appears to be a video replay of a specific play. Users can click on a dot to see a detailed view of that particular play. B) Ticketmaster Seating Selection displays a stadium seating chart on the right side. On the left is a list of available seats with prices. When a user selects a section on the map, the list updates to show seats in that section. C) ESPN Player Statistics shows a baseball field diagram in the center, surrounded by player statistics in a table format. A highlighted area suggests that certain statistics can be expanded into a dropdown for more detailed information. D) Weather.com Forecast displays a daily weather forecast with multiple time slots. Each time slot shows temperature and weather conditions. The layout is similar to the ESPN dropdown, but the caption notes that multiple detailed views can be open simultaneously.}
    
\end{figure}

Our analysis revealed variations in the arrangement of overviews and detail views, the organization of items within overviews, and the number of items displayed in detail views.

\textbf{Overview-Detail}: First, we observed variations in the layout of overviews and detail views, where some websites arranged overviews and detail views side by side, while others opened detail views in a pop up or a dropdown.

\textbf{Overview}: We also observed variations of overview layout in the forms of lists, grids, and tables. Notable variations also included visualizations that plotted points that could be opened into detail (Fig. \ref{fig:layout-dimension}A) and sports stadiums to select seating sections (Fig. \ref{fig:layout-dimension}B).

\textbf{Detail}: We finally observed that detail views varied by how many were displayed at a time. Some instances only opened one detail view at a time, while others allowed users to open many at a time. For instance, ESPN allowed only one dropdown to be open at a time (Fig. \ref{fig:layout-dimension}C), while weather.com supported multiple (Fig. \ref{fig:layout-dimension}D).

\subsection{Existing Customizability Features}
\label{section:existing-customizability}

In addition to the previously mentioned dimensions, we separately collected and coded existing customizability features for overview-detail interfaces. We found 8.33\% (13/156) of websites provided customizations for overview-detail interfaces, with 2.56\% (4/156) providing content customization, 0\% (0/156) providing composition customization, and 7.69\% (12/156) providing layout customization. All of the instances we collected were limited in customization, with the most providing eight layout variations.

We found Craigslist offered the most variations for customizing the overview-detail interface, with five options for customizing the overview layout and three for the overview-detail layout. The choice of layout also indirectly influenced which attributes were displayed. For example, the grid view included an image thumbnail, while the list view did not, likely to maintain a more compact presentation. In the content dimension, we observed four websites that allowed users to customize which content attributes they wanted to view. Among these, eBay offered the most customization options, though it only allowed users to toggle six attributes.

\subsection{Summary}
In summary, our content analysis of 303 overview-detail instances in 156 websites identified three dimensions of variation: content, composition, and layout. We found that only 8.33\% (13/156) of websites in our collection supported customization features, with those offering only limited options. Our collection of samples may not be fully representative of the entire set of information systems, and we may not have captured every instance of overview-detail interfaces within each sample. Thus, we do not claim the comprehensiveness of our design space and the frequencies of customizability features beyond our collection. Nevertheless, our analysis does provide insights into variations in real-world examples, offering contributions in understanding overview-detail interfaces.

\section{Interaction Techniques for Malleable Overview-Detail Interfaces}
\label{section:interaction_techniques}

In this section, we describe interaction techniques to customize overview-detail interfaces along the three dimensions identified from our content analysis.

\begin{figure*}
\vspace{4pt}
    \centering
    \includegraphics[width=1\linewidth]{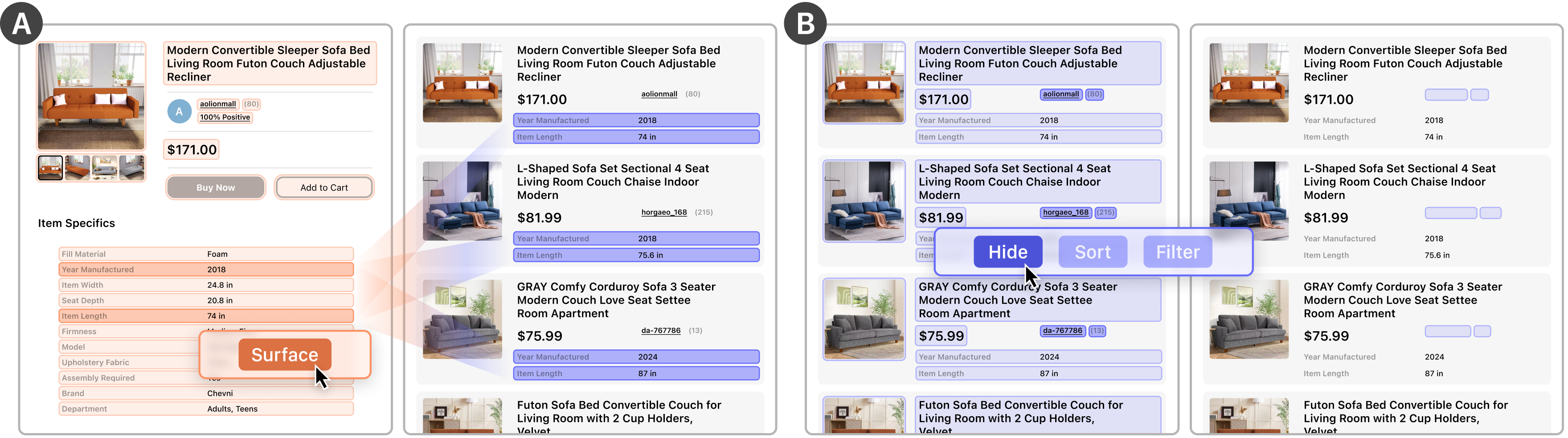}
    \caption{Upon entering ``Attributes Mode'', (A) users can select attributes in the detail view to surface to the overview, and (B) hide selected attributes from the overview. Users can modify which attributes are shown and hidden through a context menu or by directly dragging selected attributes to their desired view.}
    \label{fig:surface-and-hide}
    \Description{The figure shows five screenshots demonstrating how users might use the "Surfacing and Hiding Attributes" feature while shopping online for sofas. The first image shows a user selecting two attributes - the "Year Manufactured" and the "Item Length" - which the user surfaces. The second image shows the newly customized shopping overview - now with the selected attributes appearing alongside default attributes such as product name and price. The third image shows the user selecting the vendor's name and and hiding that attribute from the overview. The fourth image shows the final customized shopping overview - with two additional attributes surfaced and the vendor name hidden.}
    \vspace{8pt}
\end{figure*}

\subsection{Content}
\label{interaction-dimension:content}
To customize overview-detail interfaces along the content dimension, we introduce Fluid Attributes, a set of interaction techniques that allows end-users to directly interact with detail attributes in the overview-detail interface. This enables users to control what, where, and how detail attributes are shown, as well as directly sort and filter by attributes. Fluid Attributes also unlocks further customizations in overview-detail interfaces when paired with generative AI tools by allowing users to query, generate, and reformat attributes. To support such interactions, overview-detail interfaces can provide a dedicated ``Attributes Mode'', available to toggle on and off in a global toolbar on the website (Fig. \ref{fig:probe-shopping}D). Upon toggling on, the website highlights all attributes in the overview-detail interface, presenting users with affordances to manipulate and operate on the attributes.

\subsubsection{Surfacing and Hiding Attributes (Fig. \ref{fig:surface-and-hide})}
\label{feature:surface-hide-attributes}
In order to find details of items not shown in the overview, users must open the detail view of each individual item and locate the attribute in the interface. Fluid Attributes lets users select attributes from the detail view and directly surface them by clicking ``Surface'' (Fig. \ref{fig:surface-and-hide}A). The selected attributes then populate in every item of the overview. On the other hand, if users want to hide attributes they are not interested in seeing in the overview, they can instead ``Hide'' those attributes (Fig. \ref{fig:surface-and-hide}B). In essence, this interaction gives users the control to \textit{get an overview of only the things they want} by manipulating attributes on their own, without the help from the developers. 

Fluid Attributes is not limited to static attributes; it also includes dynamic and interactive elements such as widgets, buttons, and components. For instance, users can surface an image gallery component into the overview to view more images and even the \textit{Add to Wishlist} button for each item on a shopping interface.

\begin{figure}[H]
    \centering
    \includegraphics[width=1\linewidth]{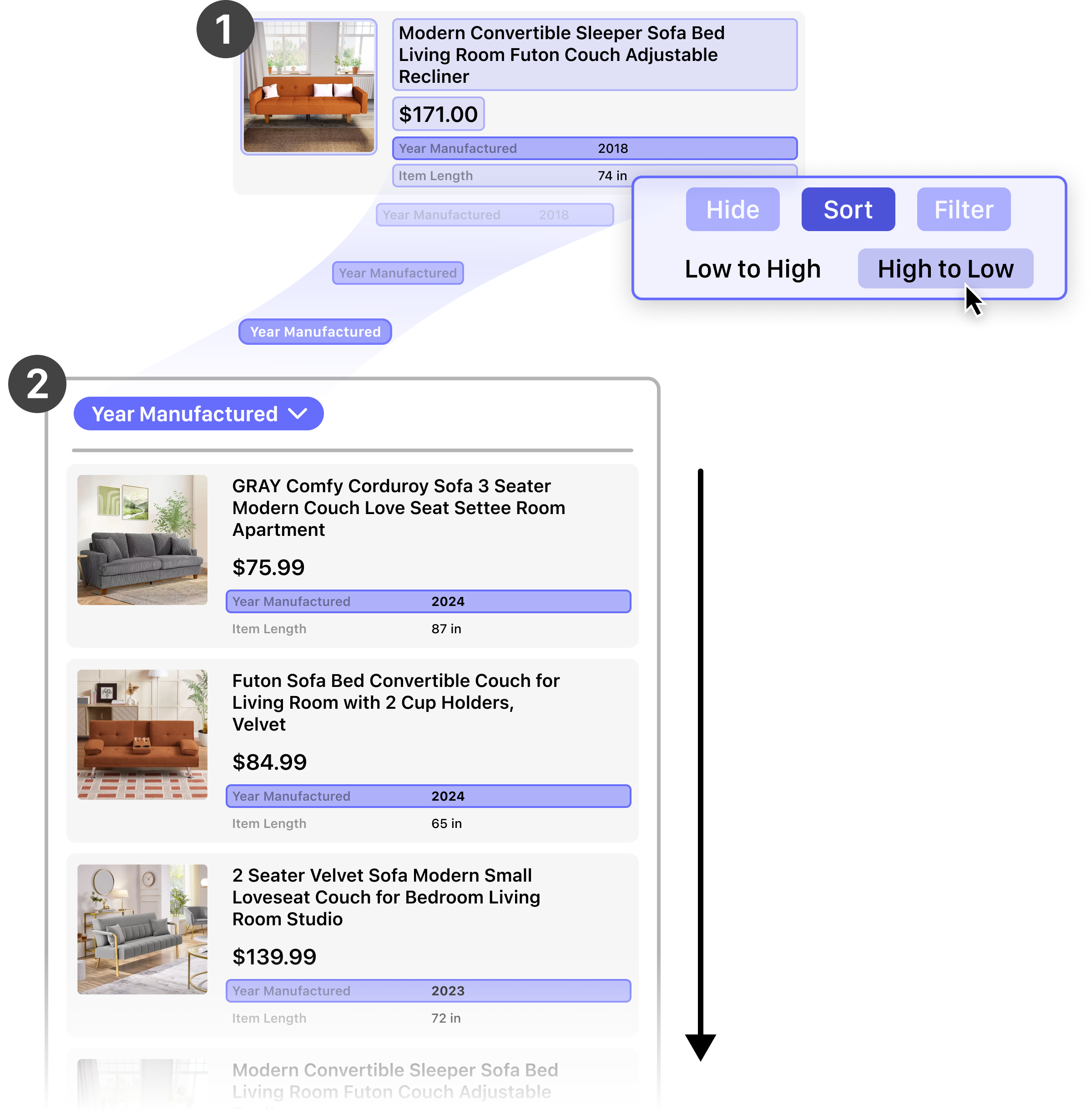}
    \caption{Since users can surface any attribute to the overview (1), this also gives them the ability to sort and filter by the surfaced attributes directly (2).}
    \label{fig:sort-and-filter}

    \Description{This figure shows how users can perform additional functions, such as sort and filter, upon any selected attribute. Since users can surface any attribute to the overview (1), this also gives them the ability to sort and filter by their values directly. Continuing with the online sofa shopping scenario, the user in the figure first surfaces the Year Manufactured attribute, and then sorts the overview items via the selected attribute.}
\end{figure}

\subsubsection{Sorting and Filtering with Attributes (Fig. \ref{fig:sort-and-filter})}

\label{feature:sort-filter-attributes}
Rather than relying on a predetermined set of sort and filter functionalities given by the interface, Fluid Attributes provides the opportunity to make any attribute sortable and filterable. For example, users can surface the \textit{Year Manufactured} attribute of a couch (Fig. \ref{fig:sort-and-filter}.1), then directly sort them by the latest manufactured year (Fig. \ref{fig:sort-and-filter}.2). Since Fluid Attributes also includes dynamic attributes, users can, for instance, sort by the boolean value associated with the \textit{Add to Wishlist} button to ``pin'' items to the top of the search results.

\subsubsection{AI-assisted Fluid Attributes}
\label{feature:ai-assisted-attributes}
Fluid Attributes, when assisted by AI's ability to process natural language queries to functionalities, demonstrates further interaction possibilities to customize the content in the overview-detail interface.
Users can leverage AI through a prompt box provided by the interface.
\begin{figure*}
    \centering
    \includegraphics[width=1\linewidth]{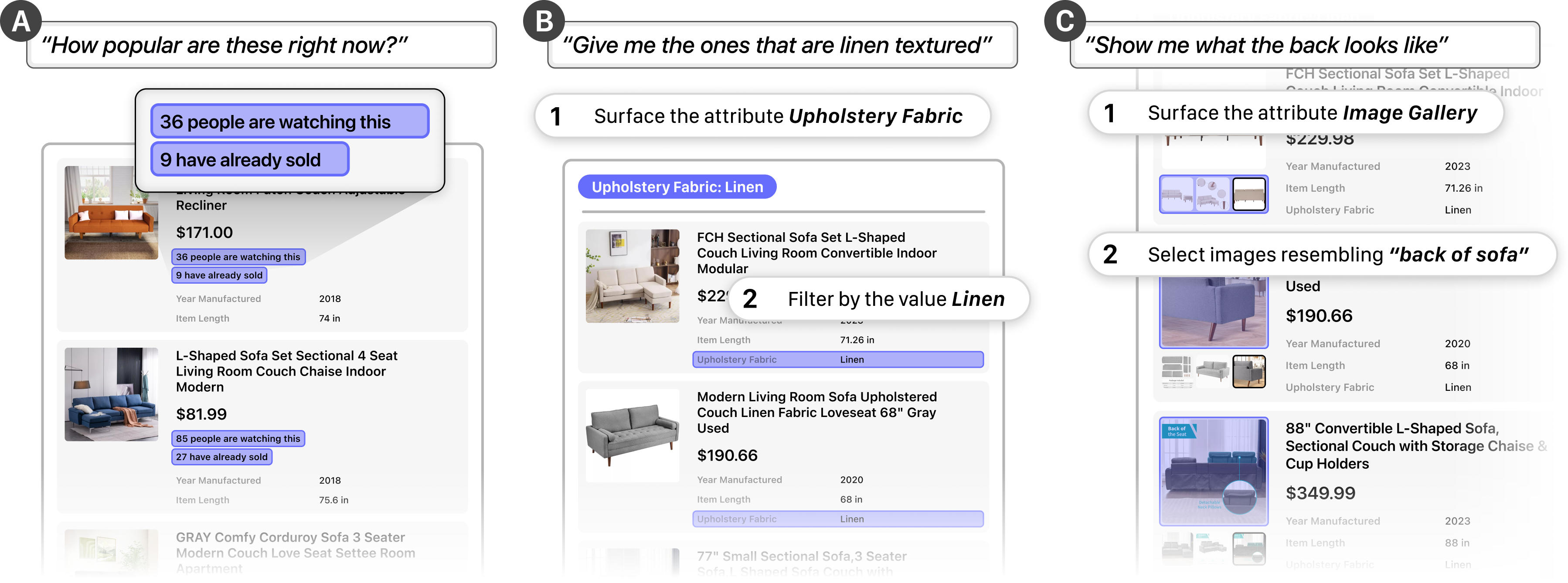}
    \caption{Users can additionally prompt AI to manipulate the attributes in the overview-detail interface. (A) Users can prompt AI to surface multiple attributes relating to the popularity of each product, (B) sort and filter for the values of a surfaced attribute, (C) and even surface specific values such as an image from the image gallery.}
    \label{fig:AI-queries}

        \Description{This figure shows three instances of how the AI feature might be used to manipulate the attributes in the overview-detail interface while shopping online for sofas. The first image shows a use case in which a user prompts "how popular are these right now" -- which surfaces prompt AI to surface attributes relating to the popularity of each product. The second image depicts a use case in which a user wants to find the sofas that are linen textured -- prompting "give me the ones that are linen textured" to surface the upholstery fabric attribute, in which the user directly filters out non-linen options. In the third image, the user prompts "show me what the back looks like", which surfaces the Image Gallery attribute. From there, the user can select images that resemble the back of the sofa.}
    \vspace{4pt}
\end{figure*}

\begin{figure*}
    \centering
    \includegraphics[width=1\linewidth]{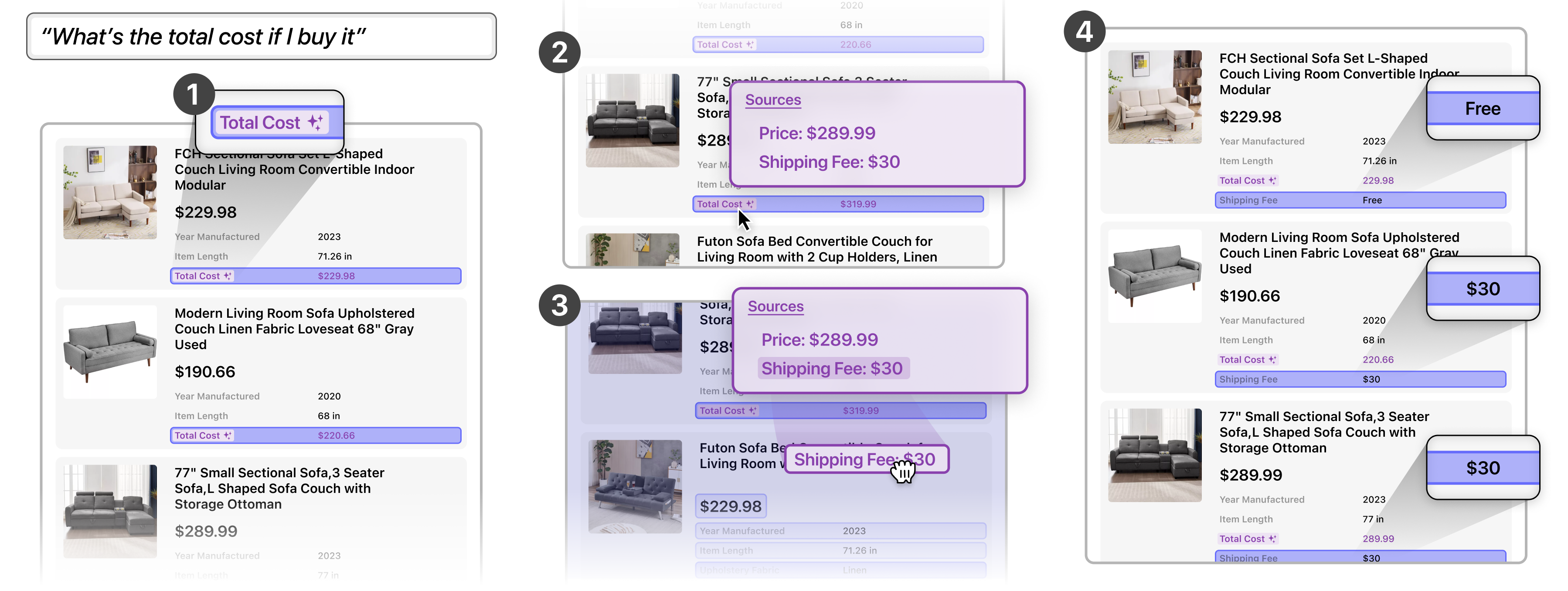}
    \caption{(1) Users can also generate new attributes with AI, for instance by asking for the total cost of the item, which sums the values of the \textit{Price} and \textit{Shipping Fee} attributes. (2) They can hover over the value to view its source attributes, and (3) also drag them into the overview, (4) surfacing the source attribute alongside the newly generated one.}
    \label{fig:AI-generated-attributes}

        \Description{This figure shows how users can generate new attributes with AI, for instance, by asking for the total cost of the item, which sums the values of the Price and Shipping Fee attributes as shown in the first image. In the second image, the user hovers over an item's Total Cost attribute to view its source attributes, and can also drag them into the overview. The final image shows the completed customization, with the newly generated attribute, total cost, alongside the source attribute, shipping fee.}
    \vspace{4pt}
\end{figure*}

\paragraph{Fluid Attributes via Natural Language Queries (Fig. \ref{fig:AI-queries})}
\label{feature:ai-query-attributes}

To customize attributes with natural language queries, users can for instance ask AI in a prompt box ``How popular are these items?'', to which AI will identify relevant attributes (i.e., \textit{Number of Watchers}, \textit{Items Sold}) and surface them to the overview (Fig. \ref{fig:AI-queries}A). This extends to the other operations in Fluid Attributes, such as adding a filter upon asking for ``only couches with a linen texture'' (Fig. \ref{fig:AI-queries}B). Users can also ask to see ``the back of the couches'', triggering AI to surface the \textit{Image Gallery} attribute and select images that contain a back view of the couch (Fig. \ref{fig:AI-queries}C). If AI fails to show this image, the user can always manually look through the images with the surfaced \textit{Image Gallery} component.

\paragraph{Generating New Attributes (Fig. \ref{fig:AI-generated-attributes})}
\label{feature:ai-new-attributes}

AI is also suited to generate new attributes from user queries. Users can ask for an attribute that may require computing values for each item in the overview.
For instance, if a user asks for the ``total cost'' of an item, AI can sum values of attributes such as \textit{Price} and \textit{Shipping Fee} to create a new \textit{Total Cost} attribute (Fig. \ref{fig:AI-generated-attributes}.1). Users can hover over the newly generated attribute to reveal the source attributes the AI referenced (Fig. \ref{fig:AI-generated-attributes}.2), and then surface these sources as well (Fig. \ref{fig:AI-generated-attributes}.3-4). If AI cannot provide the value for an item, it responds ``Not Specified''.

\begin{figure*}
    \centering
    \includegraphics[width=1\linewidth]{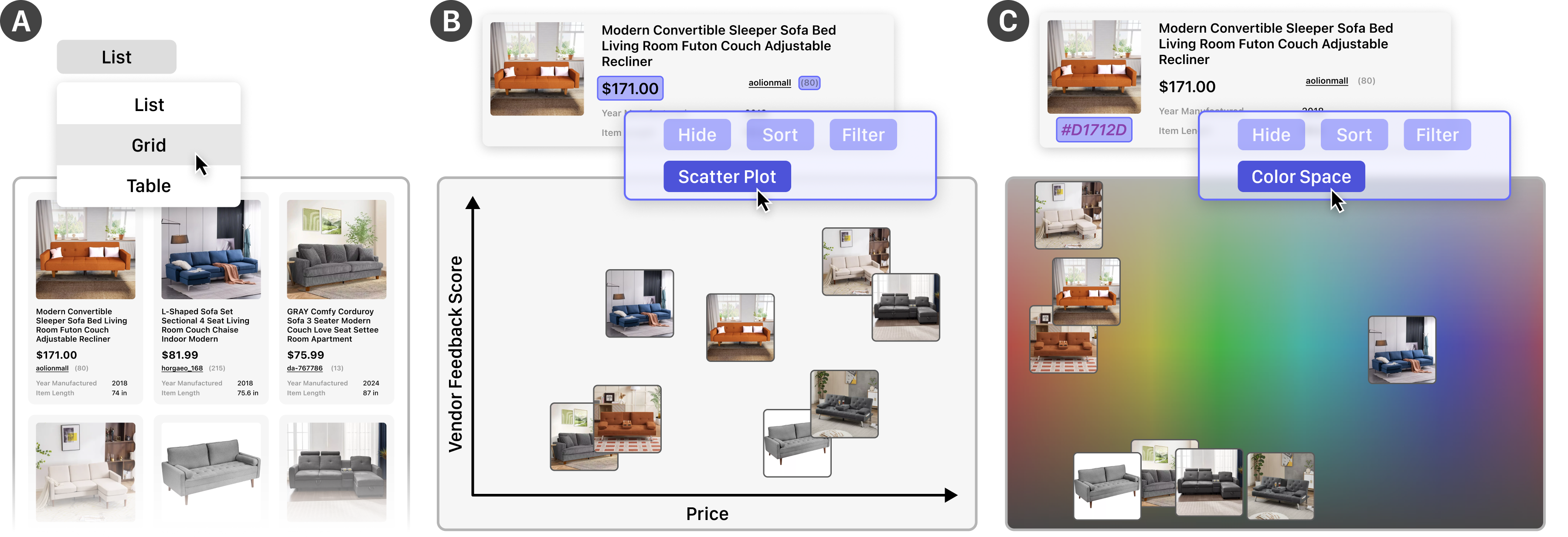}
    \caption{(A) Overview variations from the design space are available for the end-user through a dropdown menu. Fluid Attributes additionally allows users to invoke relevant overview representations through attributes. (B) For instance, users can select the \textit{Price} and \textit{Vendor Feedback Score} attributes to plot items in a scatter plot, or (C) generate a color value from the image to map items onto a color space.}
    \label{fig:overview-representations}

        \Description{This figure demonstrates how the overviews can be transformed into multiple representations with attributes. In the first image, a user selects "scatter plot" - and is able to view each overview item in a scatter plot plotted along two selected attributes as axis. The second image shows the overview items mapped onto a color space, based on a color value generated from the image.}
    \vspace{4pt}
\end{figure*}

\begin{figure}
    \centering
    \includegraphics[width=1\linewidth]{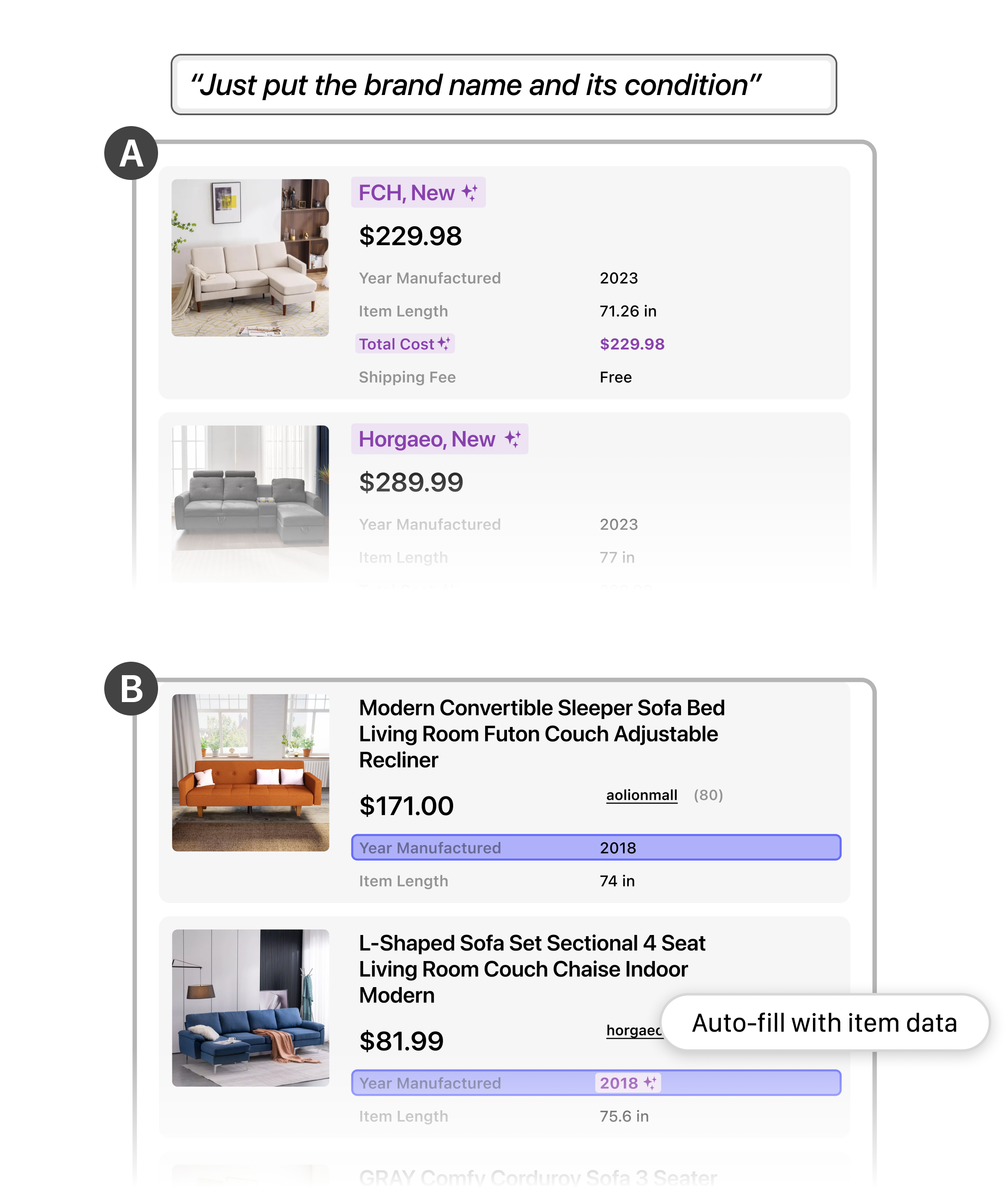}
    \caption{(A) Users can select the title and prompt AI to reformat the titles of items to make them more consistent. (B) Additionally, surfacing attributes from items that do not contain them will trigger AI to search and fill in the missing values to that attribute. If not found, AI provides the value ``Not Specified''.}
    \label{fig:AI-reformatted-attributes}

        \Description{Many attributes of items in Ebay's online marketplaces are manually entered by vendors and are often inconsistent in their formatting. This figure shows how users can prompt AI to reformat the titles of items to make more consistent -- in this case the user prompts the AI tool to "just put the brand name and its condition" -- effectively replacing the product title with the brand name attribute and condition attribute.}

\end{figure}

\paragraph{Establishing Consistent Attribute Values (Fig. \ref{fig:AI-reformatted-attributes})}
\label{feature:ai-reformat-attributes}

Many attributes of items in online marketplaces are set not only by developers, but also by vendors. This results in many inconsistent attributes and values, such as different labels, currencies, units, and title formats.
If users surface an attribute that is not found in other items, AI can scan through the item's data to identify the attribute from other sources and fill them automatically (Fig. \ref{fig:AI-reformatted-attributes}B).
Users can also reformat attributes to make them inconsistent across all items. For instance, users can select the title attribute and either ask to ``make them consistent'' or make precise changes and ask to ``show only the brand and the couch's condition'' (Fig. \ref{fig:AI-reformatted-attributes}A).

\subsection{Composition and Layout}
\label{interaction-dimension:composition-layout}

\begin{figure*}
    \centering
    \includegraphics[width=0.75\linewidth]{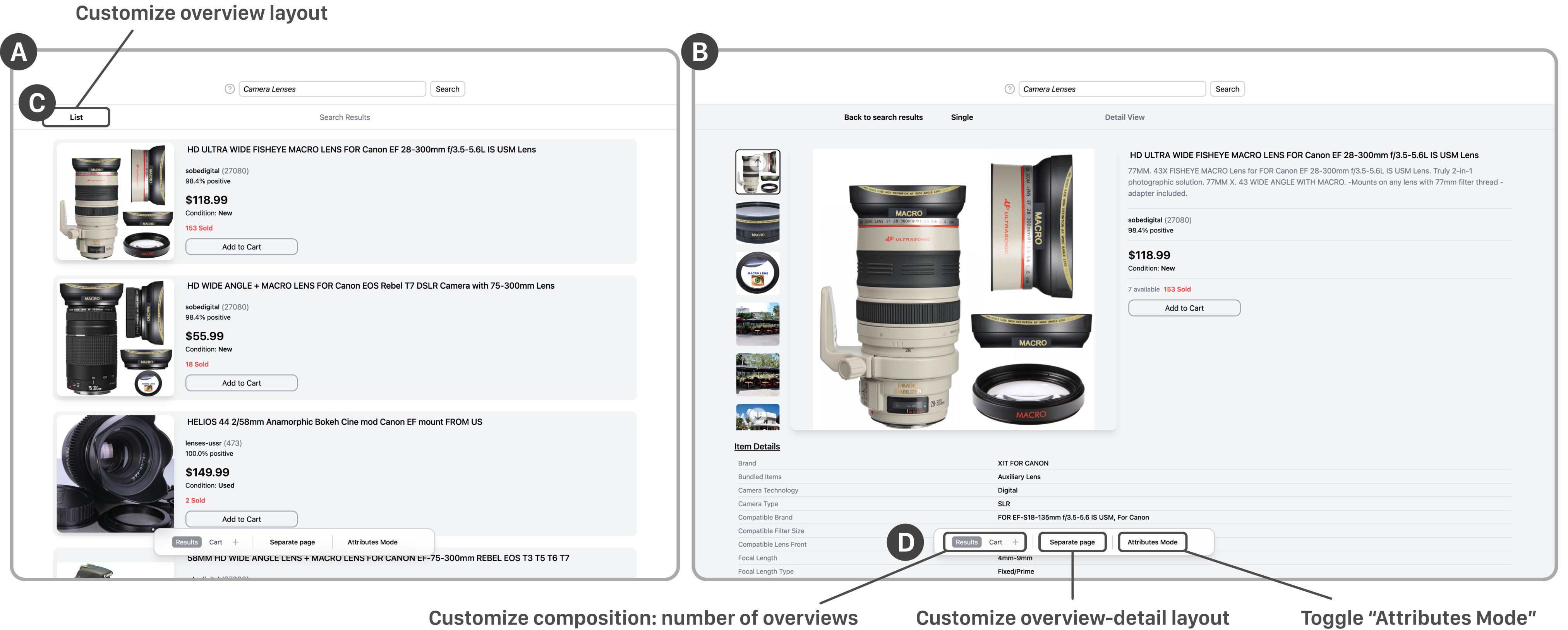}
    \caption{A screenshot of the shopping design probe, displaying (A) the overview list and (B) the detail view opened in separate pages. (C) Overview layout can be customized through a dropdown menu in the corner of each overview. (D) The toolbar provides customizations for the overview-detail, including adding and opening multiple overviews, changing overview-detail layout, and customizing attributes.}
    \label{fig:probe-shopping}

        \Description{This figure displays a screenshot of our shopping design probe. The search results overview on the left is shown as a list layout, and a product detail view is shown on the right, opened as a new page.}

\end{figure*}

To customize composition and layout, malleable overview-detail interfaces can provide variations identified by our the design space from Section \ref{section:content_analysis}.
To make transforming between such variations most familiar to end-users, we provide dropdown menus for overview and overview-detail layouts and tabbed views to present and help manage multiple overviews (Fig. \ref{fig:overview-representations}A, Fig. \ref{fig:probe-shopping}).
By implementing these familiar interaction techniques, we aim to keep the interface customizations user-friendly.

\label{feature:overview-representations}
However, other overview-detail systems may support a greater variety of overview representations to meet diverse needs, which, when listed in a dropdown menu, can become large and overwhelming for the user. We recognize that attributes are closely associated with their overview representations, as they are mapped to specific visual elements or data points within the overview.
Fluid Attributes leverages this property by enabling users to select desired properties and reveal relevant overviews in various representations.
For instance, users can select the \textit{Price} and \textit{Vendor Feedback Score} attributes to view items in a scatter plot to consider the reliability of the vendors while comparing prices (Fig. \ref{fig:overview-representations}B).
Users can also generate a color attribute from the item image to map items onto spatial overviews such as a color space, allowing users to cluster and explore items by color, similar to Histomages \cite{Chevalier2012Histomages} (Fig. \ref{fig:overview-representations}C). 
A scatter plot and color space are just two examples of overviews; other malleable overview-detail interfaces can offer additional representations and more nuanced variations of scatter plots and color spaces. Instead of aiming to provide all possible overview representations, our goal is to offer a solution that allows users to intuitively and flexibly modify how their overview is represented.

\begin{figure*}
\vspace{4pt}
    \centering
    \includegraphics[width=0.75\linewidth]{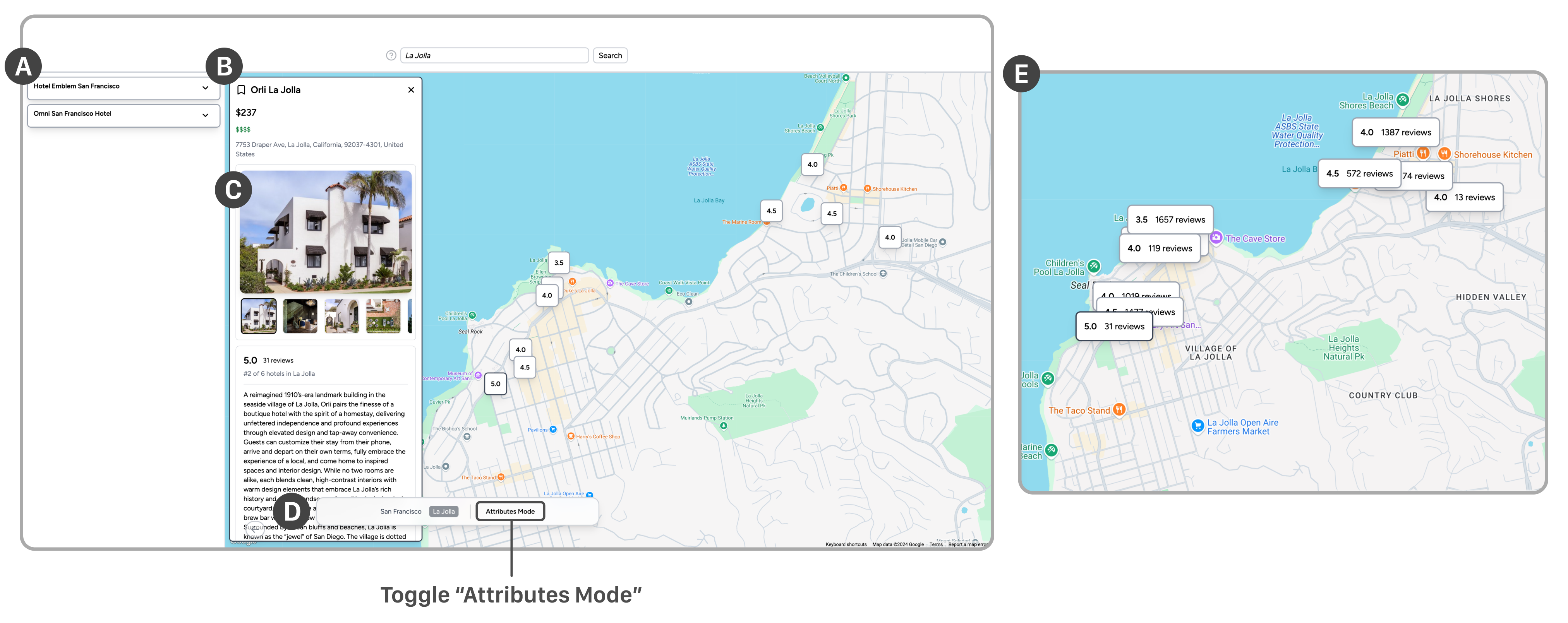}
    \caption{A screenshot of the hotel booking design probe,  displaying (A) an overview list of bookmarked hotels and (B) an overview map presenting locations as pins. (C) A detail view of a selected hotel is opened as a popover on the left side of the map. (D) The toolbar presents a way for users to jump between different searched locations and customize attributes. (E) Surfaced attributes on the map populates the pin.}
\label{fig:probe-booking}

    \Description{The figure shows a screenshot of the hotel booking design probe, displaying an overview map of La Jolla on the right and a overview list of bookmarked hotels on the left. A detail view of a selected hotel is opened as a popover left of the map.}

\end{figure*}
\section{Applications of Malleable Overview-Detail Interfaces}
\label{section:applications}

To demonstrate the generalizability of our presented customization techniques, we developed five overview-detail interfaces for various contexts and platforms: online shopping, hotel booking, managing emails, reading live articles, and ordering food on a mobile app. We developed the shopping and booking websites with higher fidelity than the others to use them as design probes for our user study. These two probes are described in more detail in Section \ref{subsection:design-probes}.

\subsection{Scenarios}

\subsubsection{Online Shopping (Fig. \ref{fig:probe-shopping})}
Online shopping websites commonly present a search result of items as a list overview with predefined attributes (e.g., item thumbnail, ``Add to Cart'' button) (Fig. \ref{fig:probe-shopping}A) and a detail view in a new page (Fig. \ref{fig:probe-shopping}B). However, this default requires users to repeatedly switch between both views to either view specific details or fully browse through individual items. A malleable overview-detail interface can instead provide a toolbar UI that presents options for the user to open detail views side-by-side or in-place with a dropdown (Fig. \ref{fig:probe-shopping}C). Additionally, users can add more overviews to save items in different collections (Fig. \ref{fig:probe-shopping}D).
This can be used to compare sets of choices such as combinations of camera bodies and camera lenses to purchase. Furthermore, users can surface attributes in the overview by toggling on ``Attributes Mode'' to directly manipulate attributes and form custom abstractions of the interface (Fig. \ref{fig:probe-shopping}D).

\subsubsection{Booking Hotels (Fig. \ref{fig:probe-booking})}
Hotel booking, home rental, and apartment search websites often provide a list view alongside a map view (Fig. \ref{fig:probe-booking}A and B) to provide rich spatial information. However, these maps often present zero or one attribute for each location, with the only way to view these details being to click on the pin one at a time (Fig. \ref{fig:probe-booking}C). Instead, a malleable overview-detail interface can enable the user to surface more details into the map, such as \textit{Rating}, \textit{Ranking}, and \textit{Amenities} (Fig. \ref{fig:probe-booking}D). Additionally, the user may customize different attributes for different overviews to leverage the benefits and affordances each overview representation provides. For instance, users can surface details about the number of reviewers in the map view to gauge which areas might be more popular (Fig. \ref{fig:probe-booking}E), while they can surface details about amenities in the list view to easily sift through preferences by scrolling.

\begin{figure}[H]
\vspace{-4pt}
    \centering
    \includegraphics[width=1\linewidth]{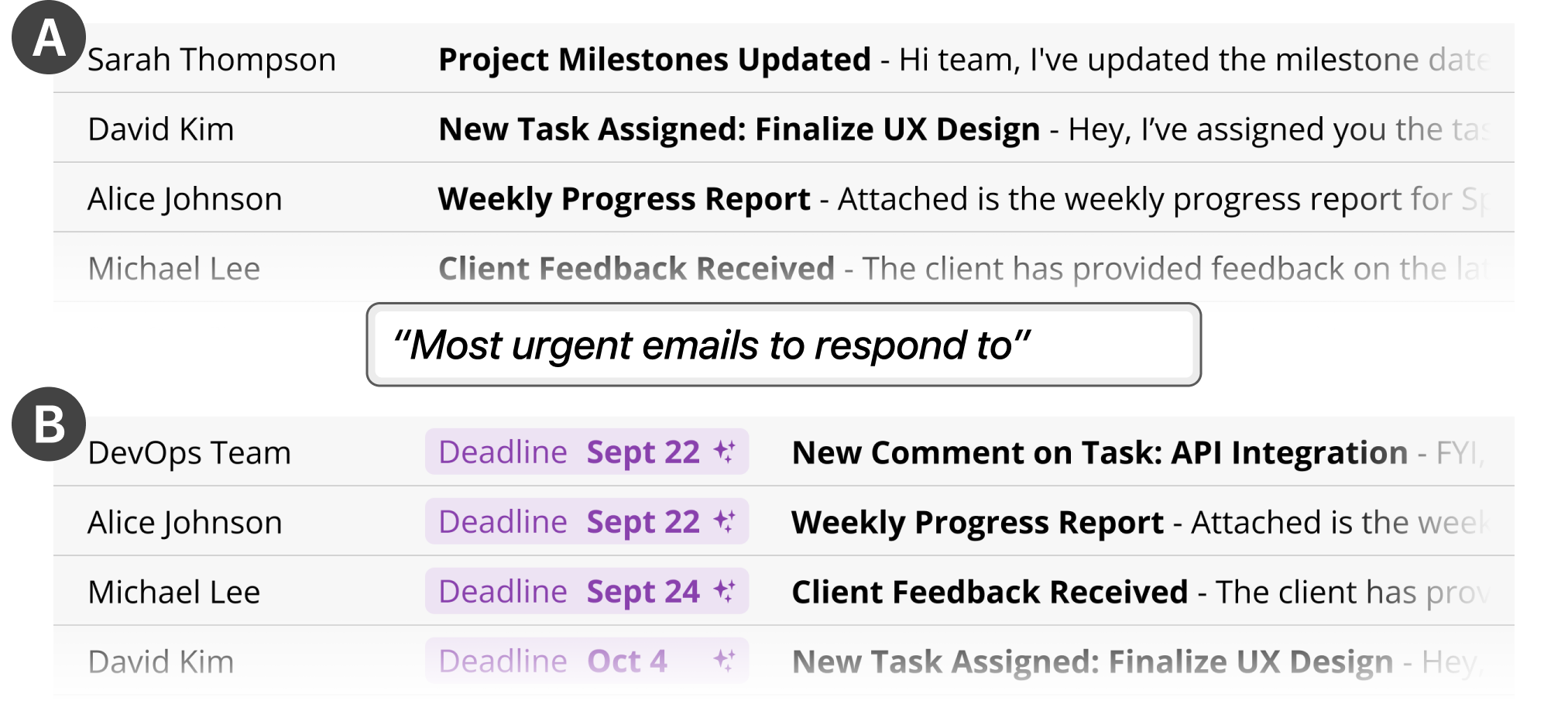}
    \caption{(A) Users can prompt AI to sort by emails with the most urgent deadlines. (B) Upon doing so, the attribute \textit{Response Deadline} is surfaced and sorted by latest date.}
    \label{fig:application-email}

        \Description{This figure depicts an email inbox as a list overview. The user prompts AI for "most urgent emails to respond to", which surfaces a newly generated attribute called Deadline that surfaces and sorts emails in the order of the response date.}
    \vspace{-4pt}
\end{figure}

\subsubsection{Reading and Managing Emails (Fig. \ref{fig:application-email})}
Oftentimes, emails contain important details in the text of their messages such as a response deadline. However, foraging through a large inbox for these details can grow extremely tedious. Using AI-assisted Fluid Attributes, users can prompt the interface to generate an attribute \textit{Response Deadline} (Fig. \ref{fig:application-email}A), surface it in the overview, and sort the emails by upcoming deadlines (Fig. \ref{fig:application-email}B).
Users can also choose to open emails in-place rather than side-by-side or on a new page, allowing them to fully read through important ones and quickly glance at less important ones.

\begin{figure}[H]
\vspace{-4pt}
    \centering
    \includegraphics[width=1\linewidth]{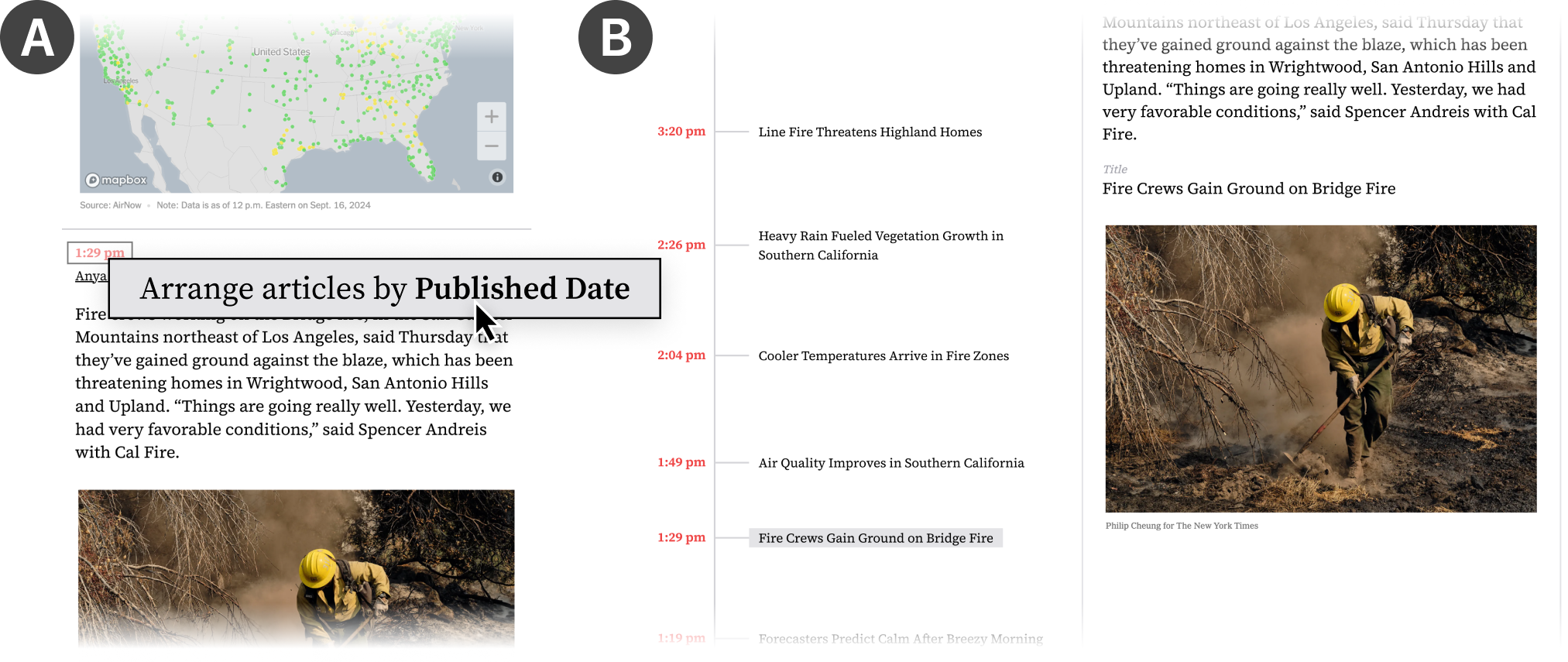}
    \caption{(A) Instead of scrolling through a long list of live updates, users can gain a broader overview of the event by selecting the \textit{Published Date} attribute and transforming the layout from a list to a timeline. (B) Users can additionally generate a \textit{Title} attribute to gain a summary of the events.}
    \label{fig:application-live-article}

        \Description{This figure demonstrates the utility of layout transformation in reading live articles. The first image (A) shows a standard scrollable stack of live articles on an online news site. The second figure (B) shows the same articles in which a user has surfaced and sorted articles via the \textit{Published at} attribute effectively transforming the layout from a list to a timeline.}
\vspace{-4pt}
\end{figure}

\subsubsection{Reading Live Articles (Fig. \ref{fig:application-live-article})}
Live news articles report on long-standing events such as presidential elections and sports games. These interfaces appear to aggregate shorter articles in a scrollable stack (Fig. \ref{fig:application-live-article}A). However, it can be challenging for users joining in the middle of the event to quickly catch up on the timeline with a long document. To address this, users can select the \textit{Published Date} attribute and bind it to a timeline overview, organizing all articles into a timeline. Users can also select which details to show in the timeline before diving into individual articles (Fig. \ref{fig:application-live-article}B).

\begin{figure}[H]
\vspace{-4pt}
    \centering
    \includegraphics[width=1\linewidth]{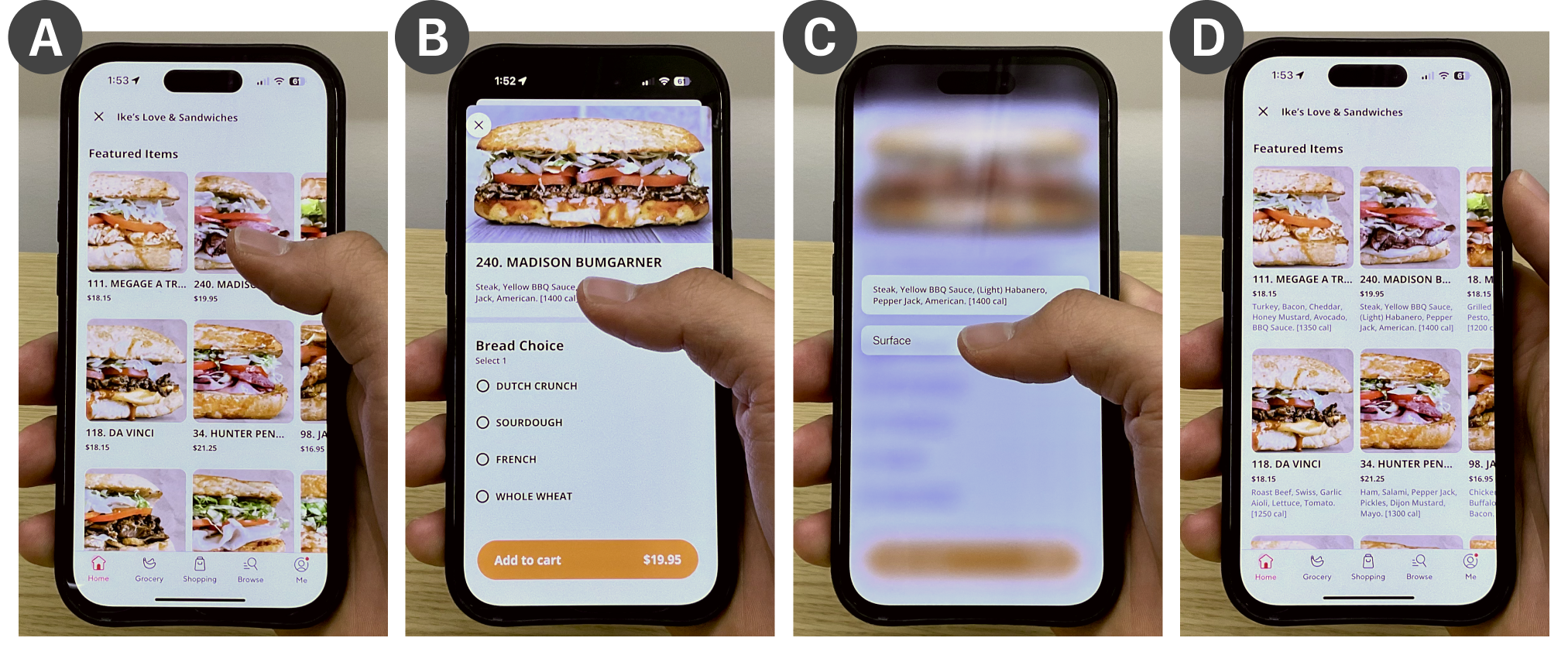}
    \caption{Our presented techniques generalize across platforms such as mobile devices. (A) Upon tapping on a menu item on a mobile delivery interface, (B) calorie-conscious users can hold down on the description text of the item and (C) tap ``Surface''. (D) This instantly reveals the ingredients and calories for every option.}
    \label{fig:application-food-order}

        \Description{This figure shows the ubiquity of manipulating attributes even on a mobile device. The image shows a standard online food delivery app, in which a health-conscious user is able to select the ingredients attribute and calorie attribute and surface them in the menu overview.}
\vspace{-4pt}
\end{figure}

\subsubsection{Mobile Applications: Food Delivery (Fig. \ref{fig:application-food-order})}
The overview-detail design pattern is also present across diverse platforms such as mobile devices. We recognize that our presented customization techniques can translate to these platforms as well.
When ordering food on a mobile delivery app, some attributes are not shown in the overview. For instance, Doordash shows prices and thumbnails of menu items to show what is popular (Fig. \ref{fig:application-food-order}A), but do not show descriptions---which includes the calories. An implementation of Fluid Attributes in this app could allow calorie-conscious individuals to check which items best suit their needs by surfacing the calorie count into the overview. 
Users can easily invoke Fluid Attributes on mobile devices by long-pressing an attribute (Fig. \ref{fig:application-food-order}B), which opens a context menu presenting available customizations, such as the ``Surface'' option (Fig. \ref{fig:application-food-order}C). Tapping this option instantly reveals the calories for all items, eliminating the need to further switch between views (Fig. \ref{fig:application-food-order}D).

\subsection{Implementation Details}
We implemented our desktop web applications in Next.js, a React web framework, while we implemented our mobile application in SwiftUI. We used eBay's API to provide results in the shopping website and used Tripadvisor's API to provide results in the hotel booking website. Our high-fidelity probes stored all customization settings on Firebase\footnote{\url{https://firebase.google.com/}. Accessed December 10, 2024.} to log them for our user study. Other applications stored customizations locally on the application's client.

\begin{figure*}
    \centering
    \includegraphics[width=1\linewidth]{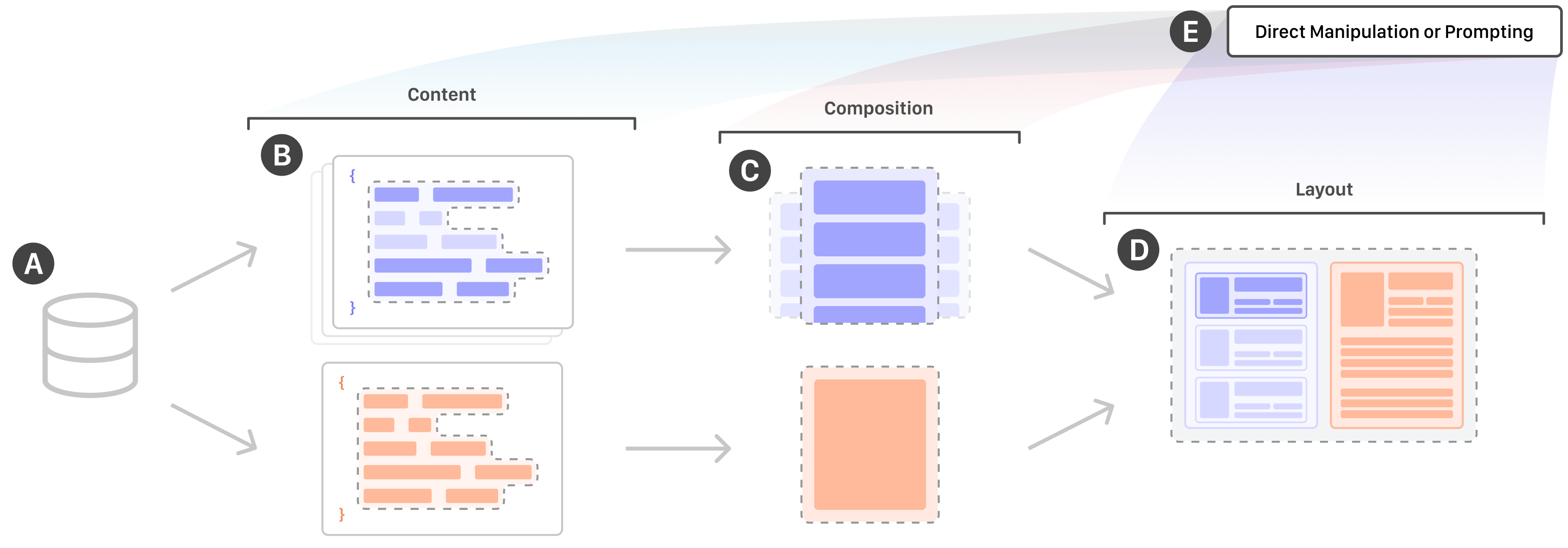}
    \caption{A system diagram of malleable overview-detail interfaces. (A) First, data for the overview and detail view is fetched from the system's database. (B) The system then determines which attributes are included in the overview and detail view. This can either be determined from fetching specific attributes from the database or from the front-end client. (C) Next, the system determines the composition of the interface, such as the number of overviews to display. (D) Finally, the set of overviews and detail views are presented in the determined layout. (E) The end-user is able to customize these three stages through the interface by either directly interacting with the settings on the interface or by prompting AI.}
    \label{fig:system-diagram}
    \Description{A system diagram of malleable overview-detail interfaces that displays stages of how data is populated to the interface. (A) contains an icon of the database, which points to the Content stage (B) which depicts JSON objects for the overview and detail views. Arrows then point to (C) which depicts views determined by composition. Finally, arrows point to (D), depicting the final interface in a particular layout. Stage (E) labels a block Direct Manipulation or Prompting which points to all three of the dimensions the user can customize.}
\end{figure*}

The pipeline of the system diagram of our malleable overview-detail implementations is illustrated and detailed in Figure \ref{fig:system-diagram}. The end-user can make customizations across the three dimensions, corresponding to the system's three components. In our applications, content customization is implemented such that all data is fetched to the front-end, and the management of visible attributes is entirely client-side. However, other implementations can opt to fetch data attributes dynamically to minimize unnecessary database calls. Both composition and layout customizations are also managed on the client side.

We implement AI to customize the content dimension only, but it can also manipulate the other two dimensions. Since all customization actions are performed through individual settings within the system, these settings can be made accessible for AI to modify based on their relevance to the user's prompt. This would also enable AI to sequence multiple customization operations together across different dimensions. The prompts used in our specific AI implementations are provided in Appendix \ref{appendix:prompts}.

\section{User Study}
\label{section:user-study}

We used our two high-fidelity applications (shopping and hotel booking) to understand how effectively users might customize malleable overview-detail interfaces and to observe emergent usage patterns as they utilize overview-detail interface customizations to best fit their needs and preferences. We aimed to answer the following questions:

\begin{enumerate}[label=\textbf{[RQ\arabic*]}]
    \item How effectively can users perform customizations on malleable overview-detail interfaces?
    \item What kinds of customization workflows and usage patterns emerge when interacting with malleable overview-detail interfaces?
    \item What additional customizations did users seek that were not supported by our design probes?
\end{enumerate}

\subsection{Design Probes}
\label{subsection:design-probes}

We selected shopping and hotel booking websites in order to approximate a common yet complex, real-world scenario of using malleable overview-detail interfaces. To focus our study on observing the use of the key customization features, we selected a subset of features that were demonstrated in Section \ref{section:interaction_techniques}.

In the shopping website's content dimension, we included surfacing, hiding, sorting, and filtering by attributes, prompting AI to surface and generate new attributes, and auto-fill attribute values.
Users can also sort and filter by attributes of each item.
For composition, we enabled users to add, remove, and rename multiple overviews. For overview layouts, we provided list, grid, and table views, and for overview-detail layouts, we provided new page, side-by-side, dropdown, and pop-up layouts.
For the booking website, we supported only the content dimension to further investigate how users surface and hide attributes and to observe potential workflow differences compared to the shopping website. Upon surfacing attributes on the booking website, users were given the option to surface them to the map, the bookmarks list, or both.
Screenshots of our probes in different configurations are in Appendix \ref{appendix:probe-screenshots}.

\subsection{Default Configurations}
We configured our overview-detail interface probes with default settings to be intuitive for participants to quickly get accustomed to. Our shopping site displayed the search results as a list overview of items. The overview-detail layout was by default a new page, and the detail view displayed one item at a time. We additionally provided a second list overview named ``Cart'', which displayed a grid layout of the items that were added to it. Each item in the overview displayed the following attributes: \textit{Title}, \textit{Thumbnail Image}, \textit{Vendor Username}, \textit{Vendor Feedback Score}, \textit{Vendor Feedback Percentage}, \textit{Product Price}, \textit{Product Condition}, \textit{Number of Products Sold}, and an \textit{Add to Cart} button. Our booking site utilized two different overviews: a map overview and an overview list of bookmarks. Upon entering the interface, the map view displayed the \textit{Rating} attribute of each hotel, and upon adding items the bookmark list, the \textit{Hotel Name} attribute was displayed by default.

\subsection{Procedure}
We conducted a 90-minute study, which began with a brief introduction of overview-detail interfaces to familiarize participants with the concepts and terminology used throughout the study (\textit{<1 min}). We then ran two 35-minute sessions, first on the booking task and second on the shopping task.
We began with the booking task because it contained a smaller feature set which could more easily acclimate users to both websites.
During each session, we first gave users a tutorial of our system, which included explanations of each feature, as well as mini-tasks to acclimate participants to our probes and discover preferred customizations (\textit{\textasciitilde 10 min}). We then proceeded onto the main task, in which participants were given a task prompt and asked to think out loud \cite{thinkaloud} (\textit{\textasciitilde20 min}). After each task, participants completed a questionnaire (7-point Likert scale) to share their experience with the probe (\textit{\textasciitilde 5 min}).

Our hotel booking task description and criteria was as follows: 
\begin{quote}
    You are traveling to the Bay Area for meetings. You will stay in two different locations, one night per stay and will need to find a hotel for each: San Francisco and San Jose. 
    \begin{itemize}
        \item Your total budget is 350 USD.
        \item You want to eat at the hotel (or hotel room) as much as possible (strong preference for hotels with included breakfast or room service).
        \item You prefer modern and new hotels over traditional or old ones.
        \item You would like to stay one night at a premium or luxurious hotel.
    \end{itemize}
\end{quote}

Our shopping task description and criteria was as follows: 
\begin{quote}
You are choosing two monitors: one for your home and one for your workplace.
    \begin{itemize}
        \item Your total budget for two monitors is 250 USD.
        \item Criteria for monitor at home:
        \begin{itemize}
            \item You want a decent deal.
            \item You want the highest resolution possible, 4k or maybe better.
            \item The monitor should take up minimum space on your desk.
        \end{itemize}
        \item Criteria for monitor at work:
        \begin{itemize}
            \item The condition of the monitor is most important.
            \item You want a display that’s decently sized, but also with high pixel density (resolution / display size).
            \item You don’t need a monitor stand---assume the office already provides one.
        \end{itemize}
    \end{itemize}
\end{quote}

At the end of the session, participants were interviewed to reflect and share their overall experience (\textit{\textasciitilde 20 min}). These interviews helped us understand their thoughts on customization preferences, strategies, limitations, and opportunities.

\subsection{Participants}
We recruited 12 participants (5 male, 7 female) from a local university through internal communication channels. Of the participants, five were graduate students and seven were undergraduate students. All participants reported having prior experience with online shopping and booking. 

Initially, we asked participants to use their own laptops for the study so it is most familiar and comfortable for them, but after four participants, we deemed users may prefer a better setup. We offered the remaining participants an external monitor (4k resolution, 31'' screen size), a keyboard, and a mouse. When offered, 7/8 participants chose to use the monitor. Participants were compensated 30 USD for 90 minutes of their time.

\subsection{Measurements}
To observe patterns in our tasks, we logged the types of customizations users made, the frequency with which they made specific customizations, and the attributes they surfaced or hid from the overviews.
To capture context and insights into participants' behaviors, we recorded both screen activity and audio.

\section{User Study Results}
\label{section:results}

First, we observed that our participants made diverse customizations even when given the same shopping and booking tasks and the same starting interface configurations, suggesting that the customizations provided by malleable overview-detail interfaces are beneficial and desired.
Many participants surfaced attributes that others had not, as illustrated in Figure \ref{fig:attributes-logs}. Some participants used multiple overviews for saving monitors into work and home lists (P1, P2, P6, P7, P12), while others created one intermediate space for candidate items (P3, P4, P8, P9, P10). While a majority of participants used the side-by-side overview-detail layout (P1, P2, P4\textasciitilde P12), P3 preferred to use the pop-up view, while P12 preferred both the dropdown and side-by-side views. Additionally, participants relied on a broad variety and combination of overview layouts between the list (P1\textasciitilde 4, P7, P9, P10), grid (P2\textasciitilde 5, P8, P10, P11), and table (P6, P11, P12). For instance, P4 used both grid and list views, while P6 used both table and list views together.

In the following sections, we report findings for each of our research questions.

\subsection{RQ1. How effectively can users perform customizations on malleable overview-detail interfaces?}

\begin{figure*}
    \centering
    \includegraphics[width=1\linewidth]{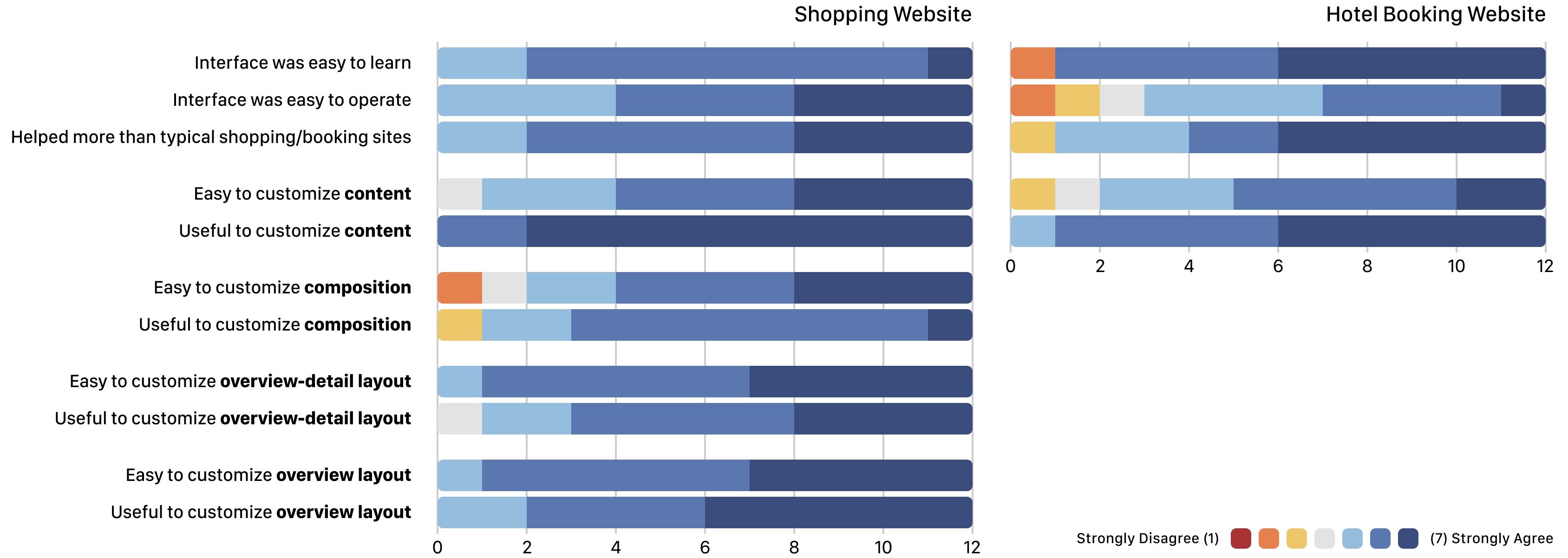}
    \caption{Participants' responses to the utility and usability of interactions to customize the content, composition, and layout the hotel booking and shopping websites.}
    \label{fig:questionnaire}

        \Description{This image shows two stacked bar charts comparing participant responses to various aspects of customization and usability for a shopping website and a hotel booking website. The chart is divided into two sections: the Shopping Website in the top half and the Hotel Booking Website in the bottom half. The Shopping Website section has 10 different criteria being evaluated, including: Easy to customize content, Useful to customize content, Easy to customize composition, Useful to customize composition, Easy to customize overview-detail layout, Useful to customize overview-detail layout, Easy to customize overview layout, Useful to customize overview layout, Interface was easy to learn, Interface was easy to operate, and Helped more than typical shopping sites. The Hotel Booking Website measures fewer criteria, only 5: Easy to customize content, Useful to customize content, Interface was easy to learn, Interface was easy to operate, and Helped more than typical hotel booking sites. For both websites, each criterion is represented by a horizontal bar, color-coded to represent different levels of agreement, ranging from "Strongly Disagree" (1) on the left to "Strongly Agree" (7) on the right. The x-axis goes from 0 to 12, representing the number of participants or the score given. In general, both websites seem to have received positive feedback, with most bars heavily weighted towards the blue (agreement) end of the spectrum. The shopping website appears to have slightly more varied responses, especially for criteria related to customizing layouts and composition.}

\end{figure*}

\begin{figure}
\vspace{4pt}
    \centering
    \includegraphics[width=1\linewidth]{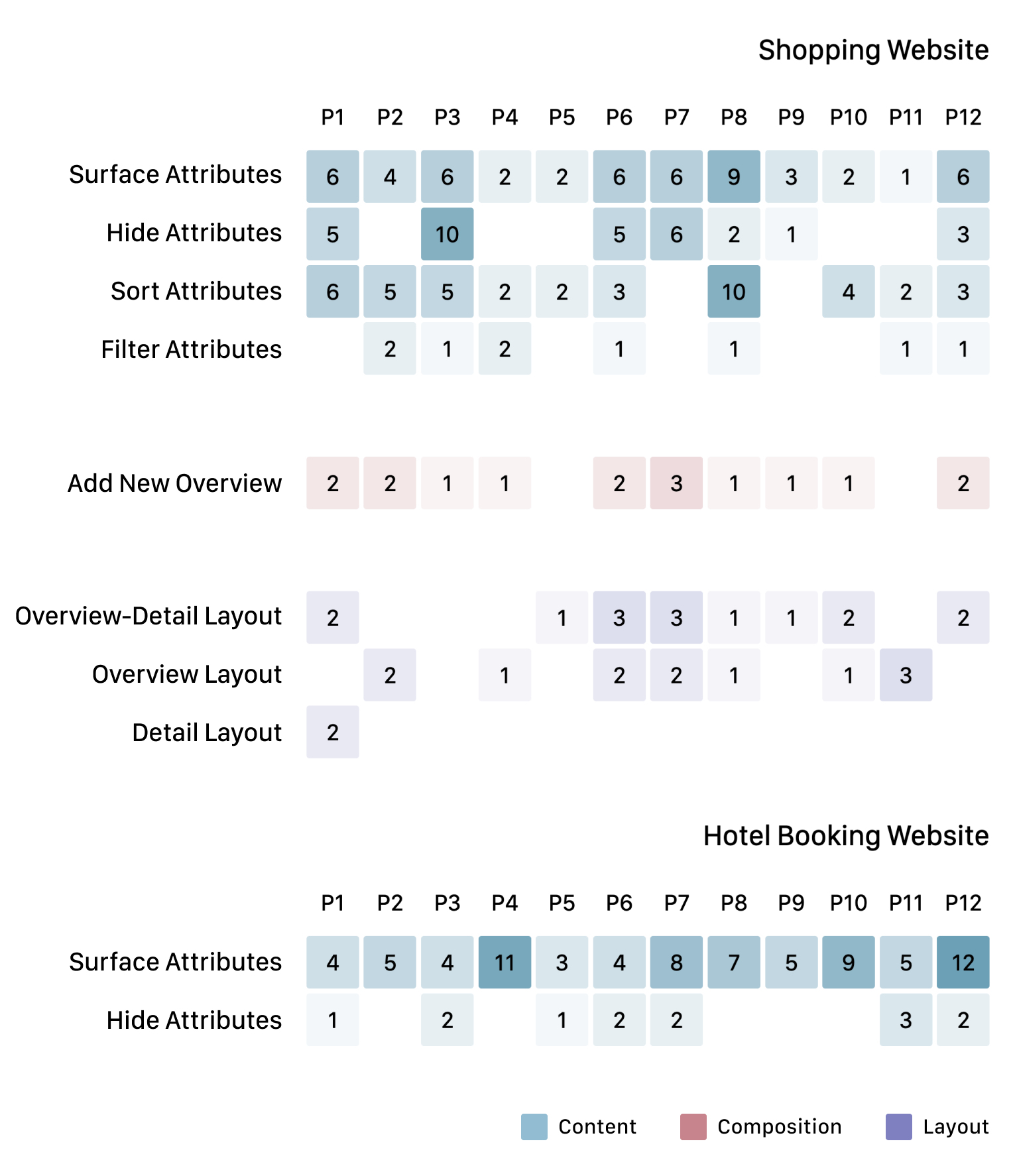}
    \caption{The number of customization operations made by each participant for each dimension. The shopping website supports customizations for all three dimensions, while the booking website supports surface and hide attributes. For the content dimension, we count how many new attributes users interacted with. For other dimensions, we count the total number of operations made.}
    \label{fig:dimensions-logs}

        \Description{This image shows a table comparing the number of customization operations made by participants (P1 to P12) for each dimension in two different websites: a Shopping Website and a Hotel Booking Website. Note that the hotel booking website supports surface and hide attributes, while the shopping website supports customizations for all three dimensions (content, composition, and layout). For the Shopping Website: the Content dimension includes Surface Attributes, Hide Attributes, Sort Attributes, and Filter Attributes. The Composition dimension includes Add New Overview, measuring how many times each participant added a new overview during their task. The Layout dimension includes Overview Detail Layout, Overview Layout, and Detail Layout. The Hotel Booking Website only includes the content dimension, which includes Surfacing and Hiding Attributes. The data includes numerical values for the frequency of each customization, but is also presented in a heat map style, with darker shades representing higher numbers of operations.}

\end{figure}

All participants completed both tasks within the provided time.
They reported that the interfaces of both probes were easy to learn (booking: 11/12; shopping: 12/12) and operate (booking: 9/12; shopping 12/12), and most agreed it was useful to customize content (booking: 6/12 strongly agree; shopping: 10/12 strongly agree), composition (11/12), and layout (overview-detail: 11/12; overview: 12/12) (Fig. \ref{fig:questionnaire}).
All participants surfaced attributes on both websites, while 9/12 participants hid attributes on at least one of them, and 10/12 participants sorted or filtered by attributes on the shopping website.
Most participants also customized the composition (10/12) and layout (11/12) of the shopping website (Fig. \ref{fig:dimensions-logs}).

We also measured how often participants made customization operations. On average, participants made 1.80 customizations per minute during the booking task (SD=0.74) and 0.86 customizations per minute during the shopping task (SD=0.48).
During their task, participants made an average of 22.08 content customizations (SD=12.29), 1.33 composition customizations (SD=0.84), and 2.17 layout customizations (SD=1.52).
These results suggest that our participants were able to effectively utilize the customizations provided to them and understood the value of malleable overview-detail interfaces. 
P1 described how the customizability features allowed them to feel more in control over the interface, saying: ``The ability to choose exactly what I wanted on my screen and close everything else was super nice. You can't really do [that] with a lot of sites. Like, it is what it is.''

\begin{figure*}
    \centering
    \includegraphics[width=1\linewidth]{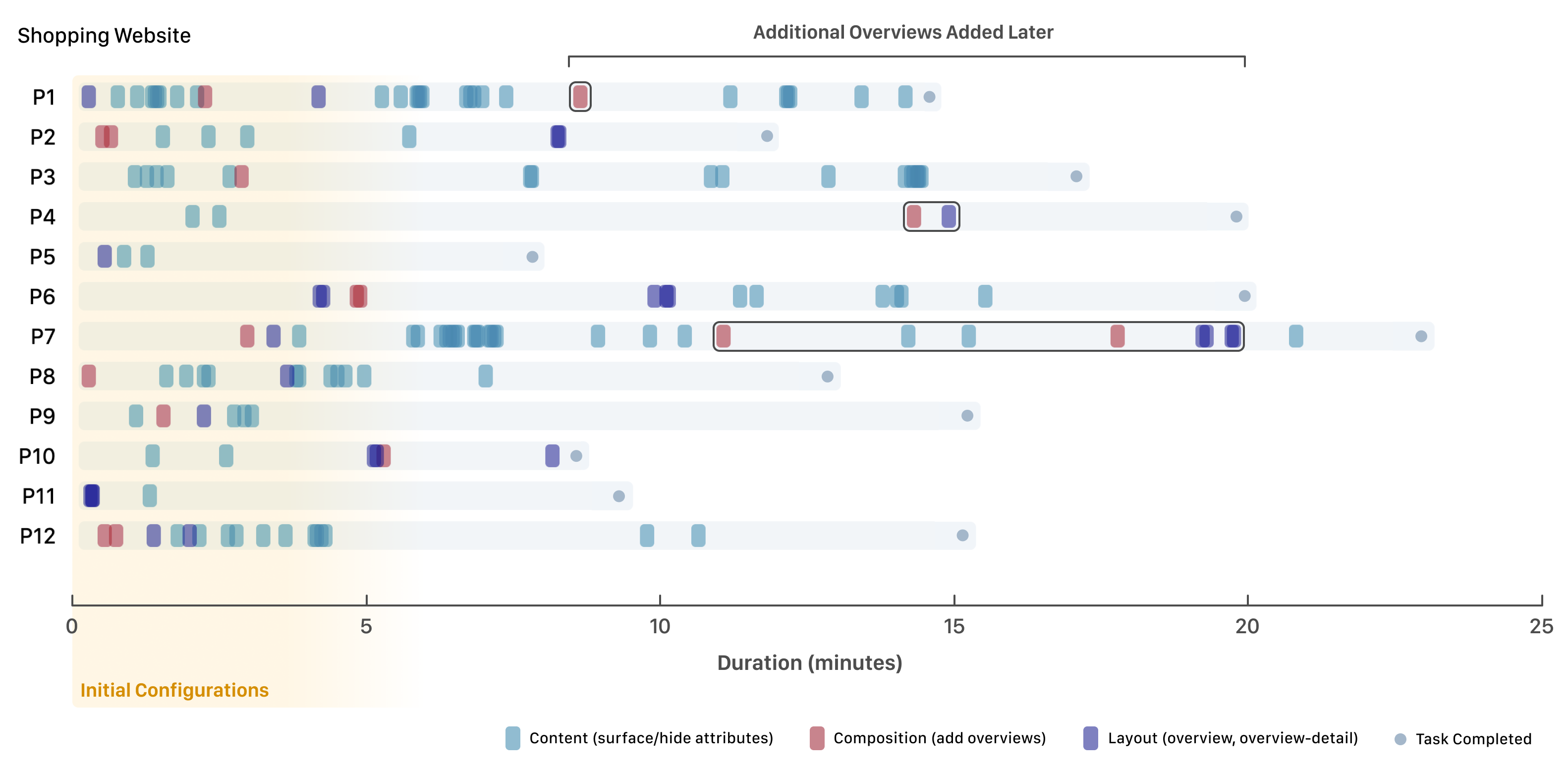}
    \caption{Logs of customizations made by participants throughout the shopping task for surfacing and hiding attributes, adding new overviews, and transforming overview and overview-detail layouts. While many participants made initial configurations, many also continued to make more throughout their decision-making processes, such as creating more overviews.}
    \label{fig:usage-shopping}
    \Description{Log visualization of each participants' customization operations throughout their shopping task. Each horizontal row is populated with content (cyan), composition (red), and layout (indigo) data points along an x-axis by duration in minutes. The figure annotates an area highlighting initial configurations for roughly the first five minutes and specific data points where users added new overviews later in the task.}
    \vspace{18pt}
\end{figure*}

\paragraph{Fluid Attributes reduces excessive context-switching}
Participants explained that the utility of customizations came from its ability to avoid unnecessary navigation by bringing all details of interest into a single space.
P3 stated, ``with a lot of websites that you can't really customize, you end up having to open the products one by one to look at all the specifications, which takes longer. I think it's nice to be able to just see it all as you're scrolling through.'' P4 similarly described their experience with the booking website compared to other commercial websites, saying, ``Websites like Booking.com, Expedia, and Airbnb don’t have these options. You have to literally look for each and everything. On the booking website, where only the options you want are displayed, it's very easy.''

\paragraph{Fluid Attributes helps users align the interface with their decision criteria}
Additionally, some participants shared that customizing content helped align the information they wanted to see on the website with their decision-making criteria. P12 said: ``After reading the information, I kind of have some criteria in my mind of which features I probably want, so I will only concentrate on these features. [With the hotel booking website], I can just customize what information I really want to focus on. For example, I personally really prefer the ratings.''

\paragraph{AI-assisted Fluid Attributes reduces the need to sift through detail}

We also observed almost all participants surfaced or generated attributes with AI (11/12). P12, who surfaced 5/6 attributes by prompting AI, found prompting the interface the most convenient as it gave them direct answers, stating: ``I'm a very lazy person, so I'll rely on this feature because I don't want to click on each [item] and check whether they are 4k monitors.'' P10 also appreciated the help from AI because they no longer felt ``confined to certain characteristics'' provided by each item. They further described how they found one shopping item that contained the \textit{Material} attribute, but noticed it did not exist for all items. Instead of manually verifying whether items are missing the attribute or have it inconsistently named, they could use AI to verify it in a single action.

\subsection{RQ2. What are the different workflows and usage patterns of malleable overview-detail interfaces?}

Findings from RQ1 demonstrate that malleable overview-detail interfaces are both effective and preferred by our participants, as the supported customizations allowed them to achieve their desired configurations. In addition to enabling these configurations, we are also concerned with how to enable users to reach these configurations more quickly. Therefore, we aim to identify the types of overview-detail configurations users prefer, which could be provided as defaults to different user groups in future interfaces to reduce the number of customization they need to perform.

\paragraph{Customizing at the start and throughout}

\begin{figure*}
    \centering
    \includegraphics[width=1\linewidth]{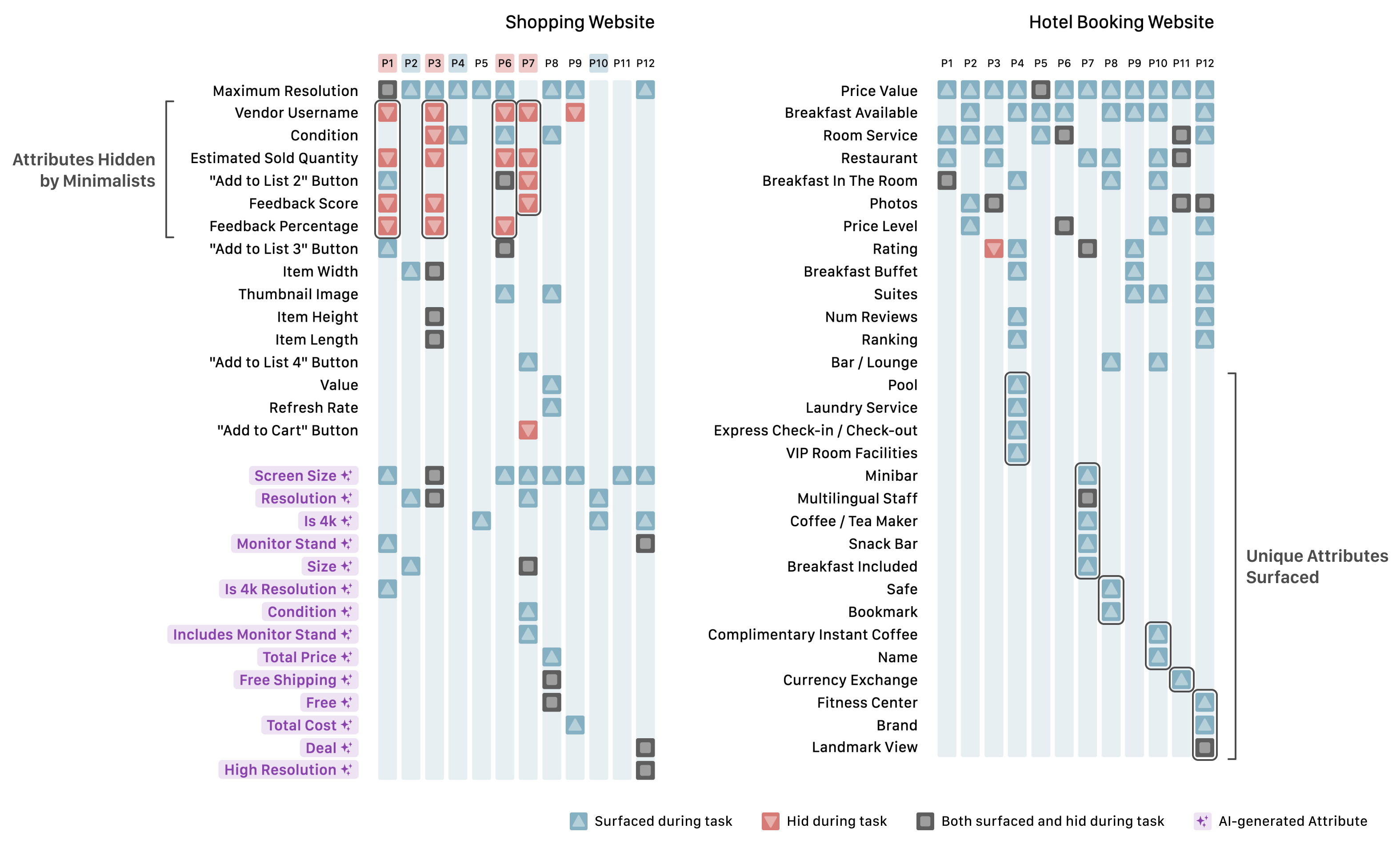}
     \vspace{-20pt}
    \caption{Visualization of all attributes that participants interacted with in the shopping and hotel booking website. The logs of the attributes that participants interacted with revealed common attributes Minimalists (\ref{results:hoarders-minimalists}) hid from the overview in the shopping website and the unique attributes some participants surfaced to the overview in the hotel booking website. We highlight participants who only surfaced attributes blue and participants who often hid attributes red.
    Attributes are sorted by the number of participants that interacted with them. AI-generated attributes are sorted separately.}
    \label{fig:attributes-logs}

        \Description{This image is a visualization comparing the various attributes that participants interacted with during the user study task: separated between the shopping website and the hotel booking website. The various attribute interactions are as follows: surfaced during task, hid during task, both surfaced and hid during task, and AI-generated attributed. The visualization is presented as a grid, with participants (P1 to P12) listed as the horizontal axis. The vertical axis consists of various attributes that participants had interacted with during the task. The Shopping website visualizes attributes such as Maximum Resolution, Vendor Username, Condition, Estimated Sold Quantity, while the Hotel Booking website visualizes attributes such as price value, breakfast available, room service, and restaurant. Each column (P1 to P12) shows how individual participants interacted with these different attributes in each respective task. The attributes axis are sorted by the number of participants who interacted with them, with the most interacted attributes at the top and the least interacted attributes at the bottom. AI-generated attributes sorted separately. This visualization ultimately allows for a comparison of attribute interactions across different participants and between the two types of websites.}
    \vspace{-8pt}
\end{figure*}

Logs of participants' customizations made throughout their shopping task are illustrated in Figure \ref{fig:usage-shopping}.
While many participants began their task by making initial configurations of their overview-detail interface (P1\textasciitilde P5, P8\textasciitilde P12), many also continued to make customizations throughout (P1\textasciitilde P4, P6\textasciitilde P8, P10, P12).
For instance, P3 initially surfaced attributes that best aligned with their criteria, but changed their set of surfaced attributes as they compared between more choices.
Additionally, some participants added multiple overviews from the start (P1\textasciitilde P3, P6\textasciitilde P9, P12), while some added them ad hoc when they needed to save and directly compare between choices (P1, P4, P7, P10).
P1, who continuously transformed their interface throughout their task, expressed the desire to stay in Attributes Mode, while P11, who preferred to ``customize once and be done'', suggested a dedicated customizations page to make their configurations in one place.
Furthermore, P6 even expressed a desire to use their table and list configuration from the shopping website for the booking website as well. They noted that, for tasks like these, they almost always default to using a spreadsheet application as their go-to tool for managing and organizing information.
To support users who continuously make customizations throughout their task, interfaces could reduce the friction of making these customization options with more direct manipulation interactions, keyboard shortcuts, or in-situ settings on the interface \cite{Ponsard2016AnchoredCustomization}.
On the other hand, those who make initial configurations could benefit from a central settings page with available customization options, popular default configurations that users can access at any point, and if possible, saved customizations across different tasks and websites.

\paragraph{Common sequences of customizations}

Most common sequences that we observed involved semantically related customizations, such as surfacing multiple attributes all relating to hotel food services or creating a new overview and immediately changing its layout.
We also observed cases when participants sorted or filtered by values of attributes they had immediately surfaced (P2\textasciitilde 4, P6, P8, P10, P11), suggesting a common \textit{surface-then-sort} and \textit{surface-then-filter} workflow.
Interfaces can reduce the number of clicks by enabling interactions for creating and using composable operations, where developers can set defaults, end-users can construct their own, and AI can generate them as well.

\paragraph{Hoarders and Minimalists}
\label{results:hoarders-minimalists}

Prior research has identified tendencies when managing digital information along the spectrum of two extremes: hoarding and minimalism \cite{Vitale2018HoardingAndMinimalism}. We found that this behavior also emerged when users were given the ability to surface or hide attributes in our probes (Fig. \ref{fig:attributes-logs}):
\begin{enumerate}
    \item Hoarders (P2, P4, P10): Participants who did not hide any attributes from the overview.
    \item Minimalists (P1, P3, P6, P7): Participants who deliberately hid attributes from the overview.
\end{enumerate}

P2 (Hoarder) did not hide any attributes in the overview, including ones they surfaced manually from the detail view or by using AI. They reasoned, ``hiding it has a bigger inertia for me to overcome because if I want it back, I would feel like I wasted more effort to remove it and bring it back again.'' They further explained that this is their natural tendency, as they also clutter their screen with many tabs. On the contrary, P3 (Minimalist), who hid five default attributes from the shopping website and one default attribute on the hotel booking website, admitted to the higher effort, but also advocated for the reward, stating, ``it probably takes longer to set it up the way I want it, but also then it's easier to look through because I'm not getting irrelevant stuff.''

Hoarders may be more resilient to clutter in the overview but less resilient to repeatedly surfacing information from the detail view. For them, an interface that packs most information in the overview, allowing them to sift through it themselves, would be preferable.
In contrast, Minimalists value the limited space in the overview and prefer to ``keep the space clean.'' For them, a blank slate in the overview, allowing them to choose which attributes to display, would be more preferable. For those who fall between these extremes, a setting to control a maximum number of attributes shown in the overview, automatically hiding less important ones when more are surfaced, may also be preferable.

\paragraph{Utilizing the overview composition and layout to structure tasks}
Among the 10/12 participants that used multiple overviews in shopping probe, five of them (P1, P2, P6, P7, P12) utilized two overviews: one for each monitor type (work and home). For instance, P12 opened only one overview at a time when finding monitors of each type to offload important information for later, saying, ``It gives me some mental space where I can preserve it.'' They also appreciated how they could open both spaces at once when they needed to compare between the two overviews. P6 did the same, but additionally split their budget into each overview, allocating \$150 for the work monitor and \$100 for the home monitor.

The behaviors of these participants hint at a pattern of utilizing the overview composition and layout to manage sub-tasks.
To support this usage pattern, future implementations of malleable overview-detail interfaces can provide support for nested or linked overviews, created either manually or with AI. This can allow users to break complex tasks into more manageable, hierarchical sub-tasks or linearly order overviews to queue sequentially dependent tasks. This aligns with literature that take the task-centric approach to provide hierarchically organized browser tabs \cite{tabsdo}, dedicated desktop spaces \cite{henderson1986rooms}, and dynamically linked email threads \cite{bellotti2003taskEmailThrasks}.

\subsection{RQ3. What additional customizations did users seek that were not supported by our design probes?}

\paragraph{Choose how detail views are invoked (P6)}
Although we implemented our probes to open detail views by clicking, P6 suggested the probes to allow users to also open details by hovering. They further described how they would also want a ``less maximized'' view of the details, suggesting the desire to invoke multiple detail views with \textit{different sets of attributes} in different ways.

\paragraph{Attributes about all items in the overview (P6, P10, P12)}
Participants wanted details across all items in the overview in addition to ones provided for each item. For instance, P6 wanted to calculate the total price of all their bookmarked hotels, while P10 wanted the ``best price'' based on multiple attributes they surfaced. 
Indeed, if the detail view can have attributes about an item, the overview can have attributes about the entire collection \cite{Xia2017CollectionObjects}.
Future implementations of malleable overview-detail interfaces should explore how this can be supported.

\section{Discussion and Future Work}
\label{section:discussion}

Our exploration of design probes and observations from our user study reveal opportunities to further investigate malleable overview-detail interfaces.
We explore future directions for supporting additional customization aspects, studying prolonged use, implementing them in real-world websites, and enabling user-defined abstractions in other contexts. Finally, we discuss how our design pattern for enhancing end-user customizability can extend beyond the overview-detail pattern.

\subsection{Supporting More Variations for Malleable Overview-Detail Interfaces}
In this research, we focused primarily on the utility of the malleability of overview-detail interfaces as a whole, rather than the utility of each individual variation. As a result, we did not implement all possible variations within the design space. However, implementing the full range of variations---and potentially more outside of our design space---could provide a more comprehensive understanding of the value each design choice offers. 
Further probes could also be implemented in contexts beyond shopping and hotel booking to offer insights into which variations and instances users might prefer under different scenarios and gain stronger evidence of various usage patterns.
Future work should explore the full range of design variations, including ones beyond our design space, and investigate their long-term utility across a variety of contexts.

\subsection{Bringing Malleable Overview-Detail Interfaces to the Real World}
Our user study showed promising signs of the utility and value of malleable overview-detail interfaces in meeting diverse user needs. 
Our work aims to serve as a concrete guide for implementing our overview-detail interface customizations across many real websites and applications.
However, the decision to implement them ultimately rests with the developers---from individual providers to large corporations. Increasing customizability requires additional development, which demands significant time, increases codebase complexity, and may even require re-architecting the application entirely. Therefore, the critical question becomes: \textit{How can we practically bring malleable overview-detail interfaces to the real world?}

We have seen many systems take the approach of foregoing the development cost entirely by creating an end-user tool as a browser extension \cite{Zhang2018Fusion, Visbug2024, Kim2022Stylette}, and this can indeed be similarly done for overview-detail interfaces by scraping active websites to identify overviews, detail views, and their associated attributes. In-fact, Sifter \cite{Huynh2006Sifter}'s DOM detection algorithm even identifies a variation of overview-detail interfaces.
Our design pattern approach could perhaps expand upon such detection algorithms to identify more variations.
However, these ``outside-in'' approaches often face challenges related to inconsistency and usability issues, making it difficult to implement them effectively.

Another approach is to reduce the cost for developers with a developer toolkit, such as a web framework, that provides many variations and customizations out of the box. This framework can package together an architecture involving GraphQL \cite{graphql2024} to dynamically fetch attributes from the back-end and a custom UI component library that provides sets of layouts and compositions to use as defaults and for end-users to customize. 
While a toolkit may be less viable for existing websites, we aim to significantly reduce the cost for developers building new ones.

\subsection{User-Defined Abstractions for All Interfaces}
Diverse forms of abstractions are implemented in various information systems and digital artifacts to assist users in understanding, operating with, and communicating with complex knowledge.
For instance, scientific data, mechanical systems, and code are often abstracted into various visualizations and diagrams, which involve the use of ellipses, blocks and arrows, and graphs \cite{ASCIIDrawings, Victor2011LadderOfAbstraction, Fong2021LinkedList}.
Recent advancements in AI have also enabled the development of generating different levels of abstractions of text \cite{Suh2023Sensecape, suh2024luminate, Graphologue} and different summarizations of academic papers \cite{elicit2024}.
We have also explored dynamic abstractions in the context of overview-detail interfaces, enabling end-users to modify a custom set of attributes.

Across these various systems and contexts, there is a clear need for end-users to define abstractions that suit their unique needs. However, the mechanisms that allow users to easily abstract information extend much beyond modifying data attributes supported in our malleable overview-detail interfaces.
For example, to create a results table from a data analysis program, users must copy the output and manually enter it into a separate formatting application, placing data into the right cells. Similarly, to share calendar availabilities to a colleague, users typically send a bullet-point list of available dates and times instead of using a scheduling tool.
In contrast, users who can define their own abstractions within the interface can generate tables directly from program outputs without manual entry and abstract away private event details from their calendar and highlight open time slots by changing the color mappings of events.
Enabling such interactions can further foster collaboration, allowing multiple colleagues to transform table and calendar interfaces to converge on a shared abstraction.
We see an opportunity to explore how users can go beyond modifying data attributes to convivially creating and tailoring their own abstractions involving semantics of text and data through AI and visual encodings with UI style. Our goal is to study user-defined abstractions as a phenomenon across these diverse contexts, seeking ways to better support them through a unified interaction methodology.
\vspace{10pt}

\subsection{Malleable Interfaces: Beyond the Overview-Detail Design Pattern}
This research focuses on making overview-detail interfaces malleable. While the overview-detail pattern is widely implemented across many interface systems, it is not the only design pattern in use. Research has explored the challenges and opportunities for end-users to customize various interface elements, including dashboards \cite{Sarikaya2019ReviewDashboards}, item collections in lists and spreadsheets \cite{Huynh2006Sifter,Litt2020Wildcard}, and menu lists \cite{Park2007AdaptableVersusAdaptive}. We have also investigated how different representations of the overview interface can be customized in relation to the overview-detail interface. We believe that exploring other types of design patterns and their design spaces can offer new opportunities for making more interfaces malleable. This approach of investigating the variations and nuances of each design pattern may help us move closer to developing a general malleable interface.
\vspace{10pt}
\section{Conclusion}
\label{section:conclusion}

This research explored malleable overview-detail interfaces, ones that end-users can customize to address individual needs. We conducted a content analysis of existing overview-detail interfaces in the wild and constructed a design space with three dimensions of variation: content, composition, and layout. Based on these dimensions, we developed interaction techniques to support customization, enabling end-users to customize diverse layouts, compositions of views, manipulable attributes across views, and AI-driven interactions to reformat and generate new attributes.
Our user study found that participants produced diverse customizations, surfacing key insights into usage patterns in how they customized overview-detail interfaces. We hope our work paves the way for further exploration of malleable overview-detail interfaces in new contexts and inspires the expansion of malleable interfaces to additional design patterns.
\vspace{10pt}



\begin{acks}
We would like to thank the anonymous reviewers for their insightful reviews to help us improve the paper.
This work was supported by NSF under grant IIS-2432644.
\vspace{10pt}
\end{acks}

\bibliographystyle{_acm/ACM-Reference-Format}
\bibliography{0_main, 0_old, 0_other}

\appendix
\clearpage
\section{Appendix}
\label{section:appendix}

\subsection{List of 33 Categories from Semrush}
\label{appendix:semrush-categories}

\begin{itemize}
    \item Advertising and Marketing
    \item Airlines
    \item Apparel and Fashion
    \item Automotive
    \item Banking
    \item Beauty and Cosmetics
    \item Computer Software and Development
    \item Computer and Video Games
    \item Distance Learning
    \item Education
    \item Entertainment
    \item Finance
    \item Food and Beverages
    \item Gambling
    \item Healthcare
    \item Hospitality
    \item Human Resources
    \item Information Technology
    \item Insurance
    \item Investment
    \item Market Research
    \item Music
    \item Newspapers
    \item Online Services
    \item Real Estate
    \item Restaurants
    \item Retail
    \item Science
    \item Sports
    \item Telecom
    \item Transportation and Logistics
    \item Travel and Tourism
    \item Wellness
\end{itemize}

\onecolumn

\subsection{Prompts Used for Design Probes}
\label{appendix:prompts}

The prompts have been slightly revised to be more presentable in the paper (i.e., renamed variables, removed conditionals inside the string, removed \texttt{JSON.stringify(...)} syntax, re-formatted syntax).

\subsubsection{Finding an Existing Attribute or Generating a New One}

\begin{verbatim}
Your job is to identify the closest attribute that exists in the provided item data given
the user prompt. This name should be an attribute of that item. If it doesn't exist, 
generate a new attribute. Do not respond with any other information.

Your response should be very short---you're just returning a single value. Wrap your 
response with brackets like so: [Answer].

The attribute name should be capitalized with normal spaces if the name includes more
than one word.

Context for this item: ${itemAttributes}

User prompt: Is it 4k?
Your response: [Is 4k]

User prompt: Multiply the item height with the length
Your response: [Item Height x Length]

User prompt: At least how old is this monitor?
Your response: [Approximate Minimum Age]

User prompt: What's the value of this monitor?
Your response: [Value]

User prompt: ${userPrompt}
Your response: 
\end{verbatim}

\subsubsection{Generating the Value of the Attribute Name}
\begin{verbatim}
Your job is to produce a value of a Attribute given the attribute name and context about 
the item.

Your response should be very short---you're just returning a single value. Wrap your 
response with brackets like so: [Answer].

If you cannot find any relevant attributes given the context provided, set the value to Not 
Specified. But for cases where the prompt is asking for existence, answer Yes or No instead.

Context for this item: ${itemAttributes}

Examples:
Attribute Name: Is 4k
Attribute Prompt: Is it 4k?
Your response: [False]

Attribute Name: Item Height x Length
Attribute Prompt: Multiply the item height with the length
Your response: [35.6]

Attribute Name: Approximate Minimum Age
Attribute Prompt: At least how old is this monitor?
Your response: [At least 2 Years]

Attribute Name: Value
Attribute Prompt: What's the value of this monitor?
Your response: [Medium Value]


Attribute Name: ${attribute.name}
Attribute Prompt: ${attribute.prompt}
Your response: 
\end{verbatim}

\subsubsection{Transforming Attribute Values}

\begin{verbatim}
Your job is to transform the value of an attribute of an item.

Your response should be very short---you're just returning a single value. Wrap your 
response with brackets like so: [Answer].

Context for this specific item: ${itemAttributes}

Example:
  Attribute to transform: ${exampleAttribute.name}
  Original attribute value: ${exampleAttribute.value}
  Transform it in this way: ${userPrompt}
  Response: ${exampleAttribute.value}


Attribute to transform: ${attribute.name}
Original attribute value: ${attribute.value}
Transform it in this way: ${userPrompt}
Response: 
\end{verbatim}
\clearpage

\subsection{Design Probe Screenshots}
\label{appendix:probe-screenshots}

\FloatBarrier

\begin{figure*}
    \vspace{-19pt}
    \centering
    \includegraphics[width=1\linewidth]{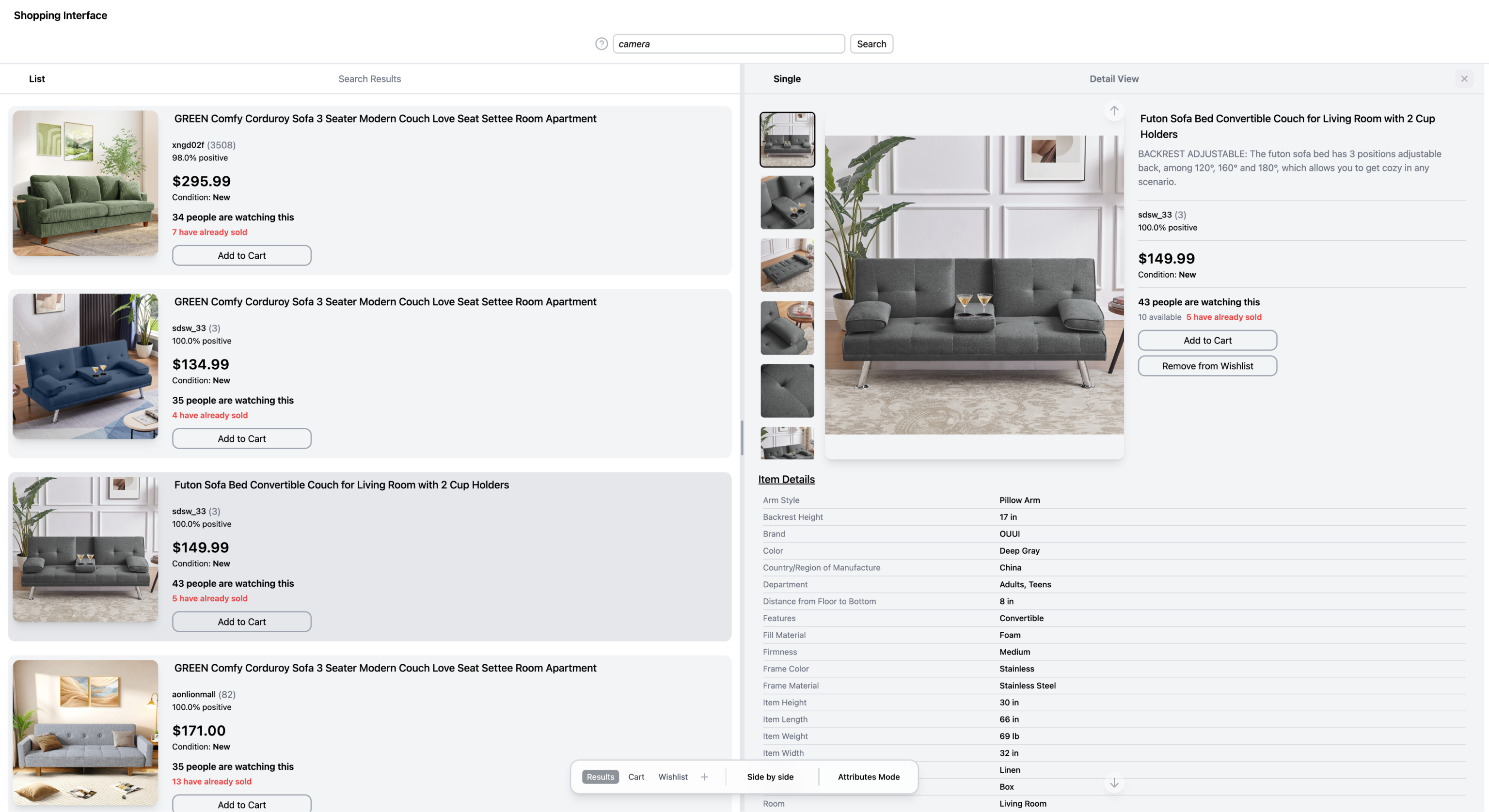}
    \caption{Screenshot of the shopping website in a side-by-side layout}
    \label{fig:appendix-shopping-1}
    \Description{Screenshot of the shopping website in a side-by-side layout}
\end{figure*}

\begin{figure*}
    \centering
    \includegraphics[width=1\linewidth]{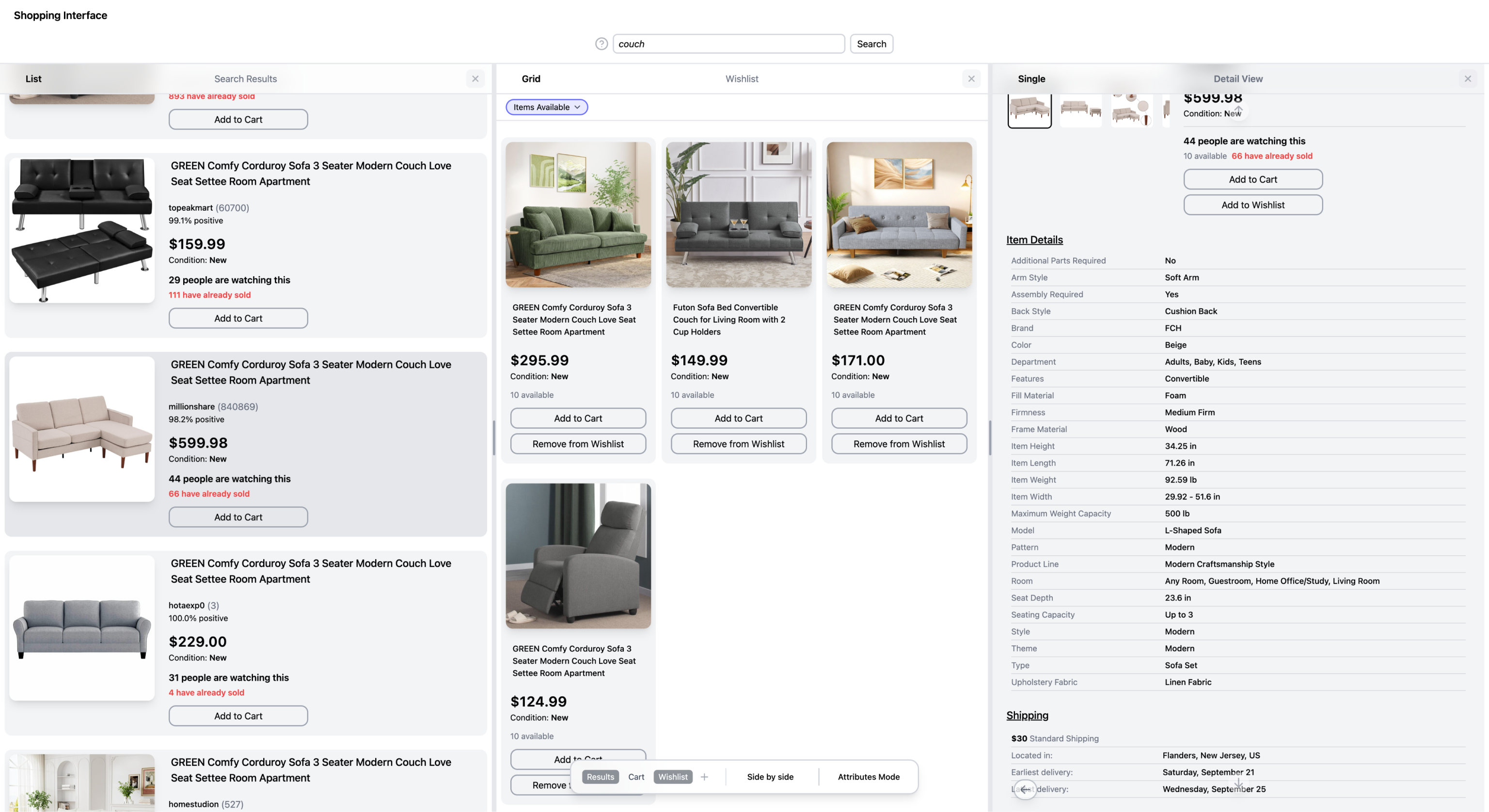}
    \caption{Screenshot of the shopping website with two overviews, one list view and one grid view.}
    \label{fig:appendix-shopping-2}
    \Description{Screenshot of the shopping website with two overviews, one list view and one grid view.}
\end{figure*}

\begin{figure*}
    \centering
    \includegraphics[width=1\linewidth]{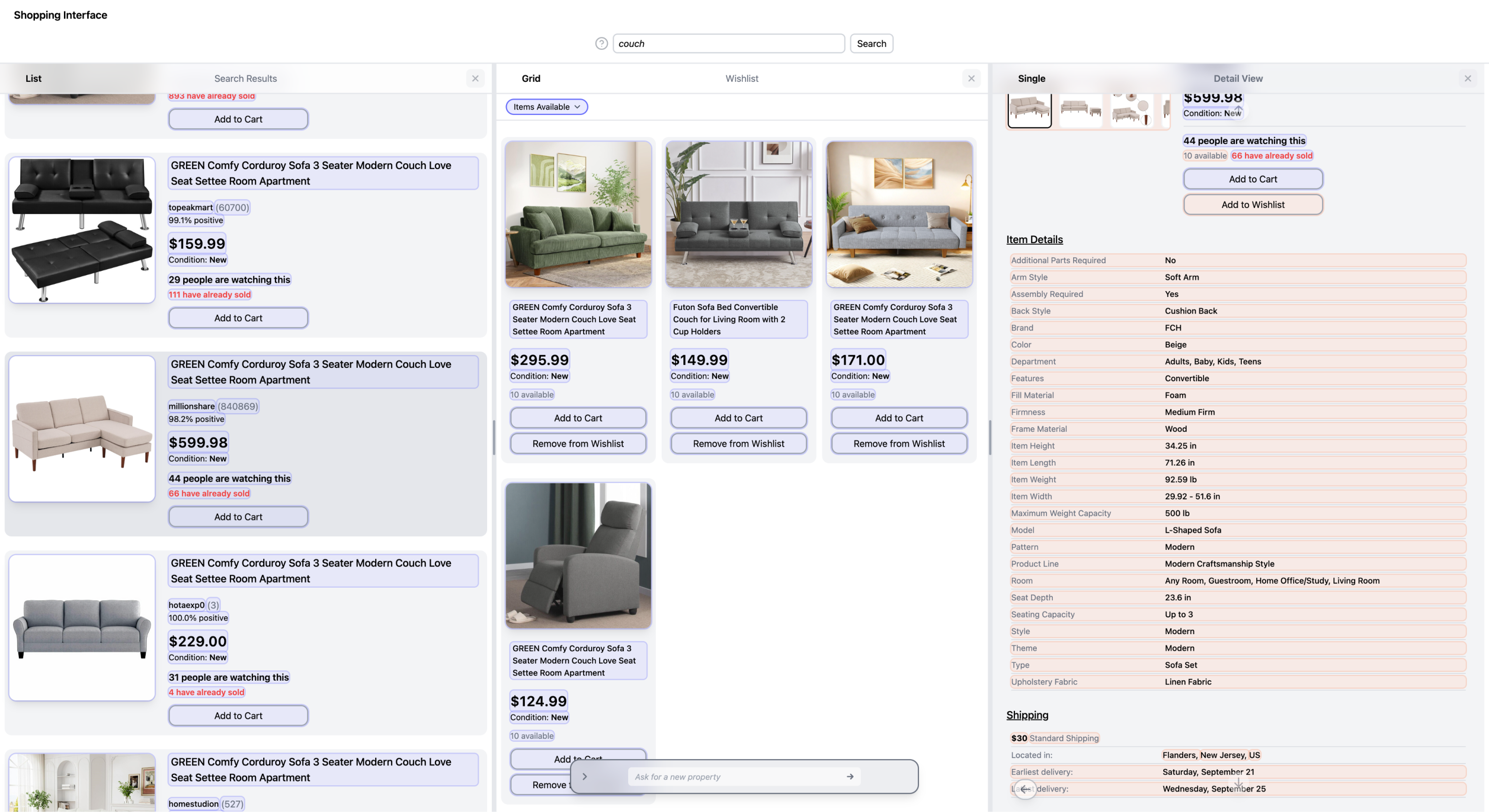}
    \caption{Screenshot of the shopping website with ``Attributes Mode'' activated. Upon activating ``Attributes Mode'', the toolbar provides a text box to prompt AI.}
    \label{fig:appendix-shopping-3}
    \Description{Screenshot of shopping website with ``Attributes Mode'' activated. Upon activating ``Attributes Mode'', the toolbar provides a text box to prompt AI.}
\end{figure*}

\begin{figure*}
    \centering
    \includegraphics[width=1\linewidth]{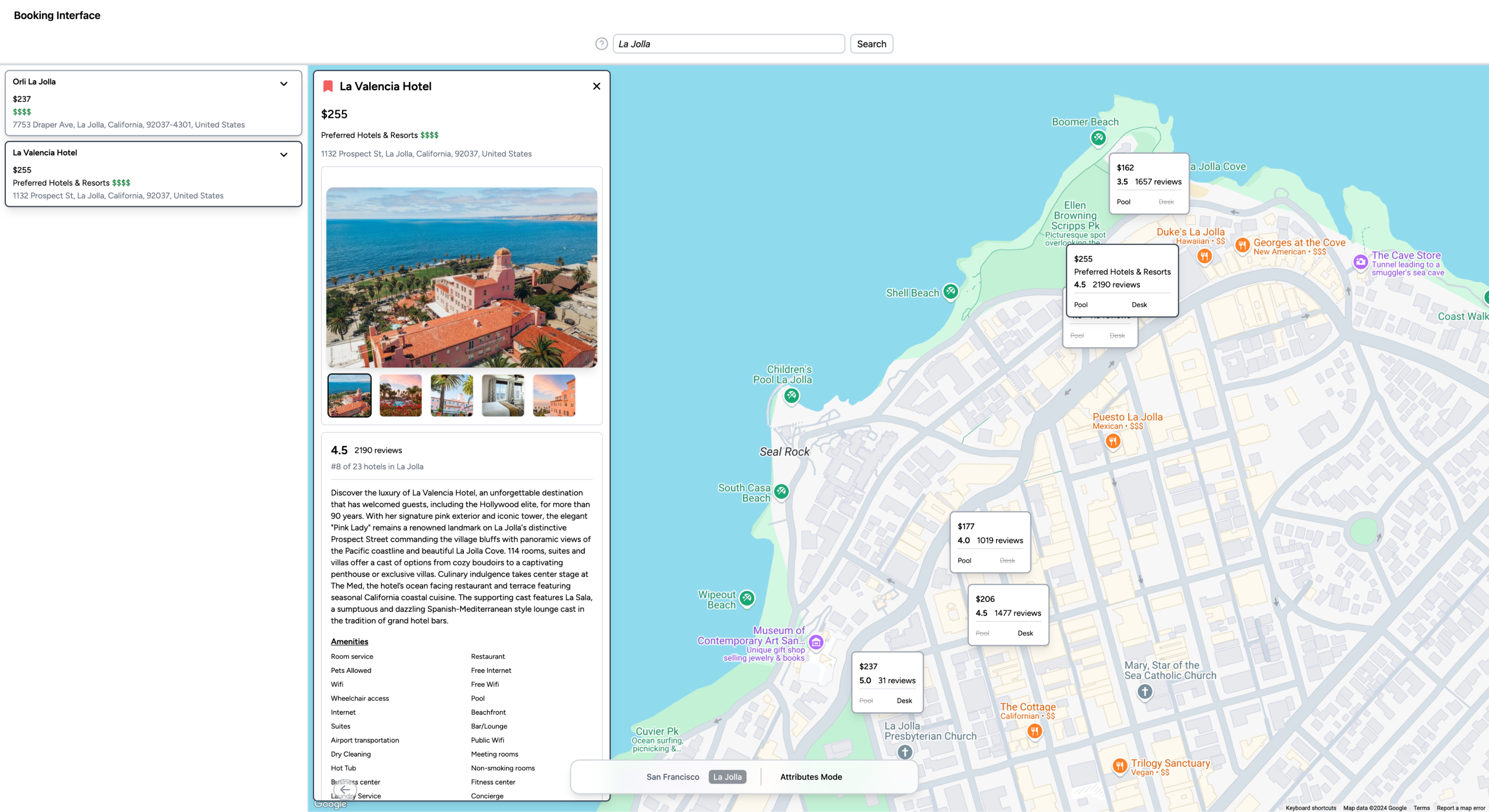}
    \caption{Screenshot of the booking website with different attributes surfaced in the bookmarks list and map view.}
    \label{fig:appendix-booking-1}
    \Description{Screenshot of the booking website with different attributes surfaced in the bookmarks list and map view.}
\end{figure*}

\end{document}